\documentclass[prl,amsfonts,amssymb,floats,twocolumn,superscriptaddress,aps]{revtex4-2}
\usepackage{amsmath,amsfonts}
\usepackage{bbm}
\usepackage{float}
\usepackage[caption = false]{subfig}
\usepackage{graphicx}
\usepackage{color}
\usepackage{hyperref}
\usepackage[normalem]{ulem}
\newcommand{\w}{\omega}
\newcommand{\jp}{J}
\newcommand{\jpl}{K}
\newcommand{\sv}{\vec{S}}

\newcommand{\msg}{m_{\rm sg}}

\newcommand{\Kc}{K_{\rm c0}}

\newcommand{\Ztwo}{\mathbb{Z}_2}

\newcommand{\llangle}{\langle\!\langle}
\newcommand{\rrangle}{\rangle\!\rangle}

\graphicspath{{./}{./images/}}

\usepackage{pdfpages}
\usepackage{pgffor}
\makeatletter
\AtBeginDocument{\let\LS@rot\@undefined}
\makeatother

\begin{document}

\title{Quantum spin-glass criticality in disordered frustrated dimer magnets}
\date{\today}

\author{Darshan G. Joshi}
\affiliation{Tata Institute of Fundamental Research, Hyderabad 500046, India}
\author{Matthias Vojta}
\affiliation{Institut f\"ur Theoretische Physik and W\"urzburg-Dresden Cluster of Excellence ctd.qmat, Technische Universit\"at Dresden, 01062 Dresden, Germany}

\begin{abstract}
We study quantum phase transitions of Mott insulators between spin-glass and featureless paramagnetic phases. Specifically, we consider a triangular-lattice bilayer Heisenberg model with bond disorder. The clean system has two phases, a dimer quantum paramagnet and a non-collinear antiferromagnet, separated by a quantum critical point. Bond disorder destroys the antiferromagnetic phase via the interference of dipolar textures and results in spin-glass order, such that the system features a quantum phase transition between spin-glass and dimer phases. We study the vicinity of this transition using a variant of bond-operator theory and calculate thermodynamic observables as well as excitation spectra. Bond disorder leads to strong inhomogeneities near the transition which suppresses non-collinearities and leads to anomalously weak glassiness in the near-critical quantum spin glass. We characterize the low-energy excitations which have a strong tendency towards spatial localization, and we track the behavior of the amplitude (i.e. Higgs) mode across the glass phase whose spectral weight we find to be strongly suppressed due to interference effects.
\end{abstract}

\maketitle

%%%%%%%%%%%%%%%%%%%%%%%%%%%%%%%%%%%%%%%%%%%%%%%%%%%%%%%%%%%%%%%%%%%%%%%

The combination of frustration and quenched disorder in magnetic insulators often leads to spin-glass behavior \cite{villain79,fischer91,parisi_book}. Spin glasses constitute a fascinating and, at the same time, notoriously difficult area of research: In addition to non-trivial thermodynamic properties, they are characterized by complex hierarchical dynamics. Many theoretical studies in the past have focussed on simple classical models, such as the Sherrington-Kirkpatrick \cite{sherrington75}, the Edwards-Anderson \cite{edwards75} model, and their descendants, with a key question being the existence and nature of a thermal phase transition into the glass state beyond mean field \cite{parisi_book,mezard84,franz94,baity13,maiorano18,angelini22}.

A natural extension are quantum spin glasses and their phase transitions, adding complexity arising from quantum fluctuations and entanglement \cite{georges00,christos22,chowdhury_rmp,kavokine24,viteritti25}. Conceptually, quantum phase transitions (QPT) out of spin glasses can occur towards different phases, with at least the following possibilities \cite{rop_fru}: (i) to a phase with magnetic long-range order, (ii) to a fractionalized spin-liquid phase, (iii) to a disorder-dominated paramagnet, such as a valence-bond glass or random-singlet state, and (iv) to a featureless and asymptotically homogeneous paramagnet. All of these bear experimental relevance, given that frustrated magnets can be tuned by pressure, magnetic field, and controlled chemical substitution. Clearly, the cases (i--iv) will lead to very different transition properties and therefore require different theoretical treatments; the focus of the present paper will be on case (iv).

Given the difficulties in understanding the low-energy physics of spin glasses in general, there is relatively little work on spin-glass quantum criticality. One large strand of activities, also belonging to case (iv) above, deals with Ising spin glasses in a transverse field, with early work on the infinite-range (or all-to-all) model \cite{bray80,huse93} and subsequent papers on finite-range lattice models in $d=2,3$ \cite{rieger94,miyazaki13,matoz16}. These transitions, also relevant for fundamentals of quantum annealing, remain incompletely understood \cite{bernaschi24}.
Much less work has been done for situations with continuous symmetry, mainly restricted to all-to-all random Heisenberg models \cite{georges00,georges01,biroli02,arrachea02,shackleton21,christos22}.
Studies for realistic lattice models are largely lacking, the recent works in Ref.~\onlinecite{viteritti25,bracci26} being notable exceptions.

\begin{figure}[!tb]
\includegraphics[width=\columnwidth]{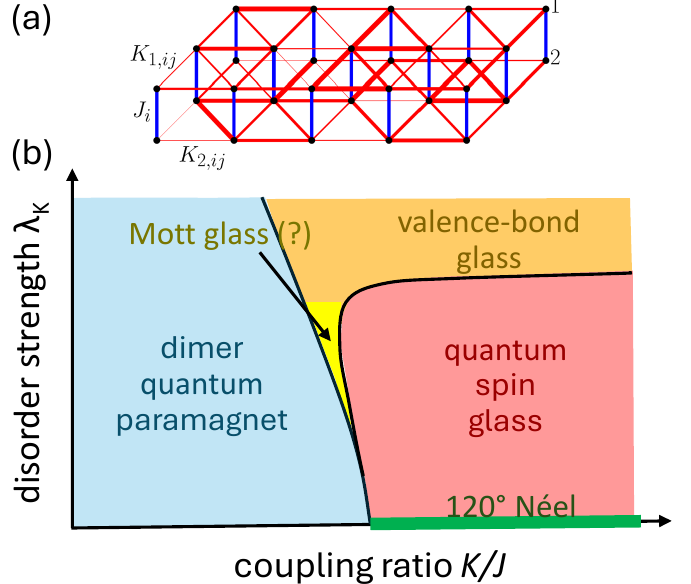}
\caption{
(a) Bilayer triangular lattice Heisenberg model, shown with disorder in the intralayer bonds.
(b) Qualitative phase diagram as function of the coupling ratio $K/J$ and the disorder strength $\lambda_K=\Delta K/K$, for details see text.
}
\label{fig:phd}
\end{figure}

In this paper we aim at partially filling this gap: We study a quantum Heisenberg model with frustration and quenched disorder which can be tuned between multiple phases. We focus on the zero-temperature transition between a dimer paramagnet and a spin glass with non-collinear spin correlations; this transition can be accessed using a variant of dimer parton (or bond-operator) theory. We compute static and dynamic observables and characterize both the states and the energy landscape in the glass phase: We find that glassiness is doubly weak near the transition, as a result of both the moment amplitudes and the non-collinearities vanishing at the transition. The glassy state is non-coplanar, such that transverse and longitudinal excitation always mix. We compute the dynamic scalar (i.e. singlet) susceptibility to study the amplitude (Higgs) mode near the quantum phase transition.
We provide predictions for the dynamic spin structure factor and discuss connections to experiments.

%%%%%%%%%%%%%%%%%%%%%%%%%%%%%%%%%%%%%%%%%%%%%%%%%%%%%%%%%%%%%%%%%%%%%%%

\paragraph{Model.--}
We consider a Heisenberg model on a triangular lattice bilayer, Fig.~\ref{fig:phd}(a), with the Hamiltonian
\begin{equation}
\label{eq:Ham}
\mathcal{H} = \sum_{i} \jp_{i} \sv_{1i} \cdot \sv_{2i}
+ \sum_{\langle ij \rangle,m} \jpl_{m,ij}  \sv_{mi}\cdot\sv_{mj} \,,
\end{equation}
where $\sv_{mi}$ represents a spin $1/2$ on layer $m=1,2$ and dimer site $i$, and $\jp_{i}$ ($\jpl_{m,ij}$) are the antiferromagnetic interlayer (intralayer) couplings, respectively.

We quickly discuss the model's phases without quenched disorder, with parameters $\jp\equiv\jp_{i}$ and $\jpl\equiv\jpl_{m,ij}$; we also introduce a second-neighbor intralayer coupling $\jpl'$. An interlayer singlet phase without broken symmetries, i.e., a featureless paramagnet, occurs for $\jp\gg\jpl,\jpl'$. In the opposite limit of weakly coupled layers, the model realizes bilayer versions of the phases known from the single-layer triangular-lattice Heisenberg model, with 120$^\circ$ spiral order for $\jpl'/\jpl<0.08$, four-sublattice stripe order for $\jpl'/\jpl>0.16$, and one or more quantum spin-liquid phases in between \cite{jolicoeur90,zhuwhite15,iqbal16,eggert19,ferrari19,tang22}.
Quantitatively accurate studies of the phase boundaries in the bilayer case are scarce: For $\jpl'=0$ the transition between the dimer and 120$^\circ$ phases has been estimated from series expansions as $\jpl/\jp\approx 1.2$ \cite{singh98}; the transition is in the chiral universality class of the stacked triangular antiferromagnet \cite{kawamura92}. We note that bond-operator mean-field theory delivered a very different value $\jpl/\jp\approx 6.6$ \cite{gu00}; this discrepancy indicates that this mean-field theory strongly underestimates the stability of the ordered phase. In contrast, the linearized bond-operator approach to be employed below yields a transition point at $\jpl/\jp=1/3$; the inclusion of interaction corrections via the Brueckner approximation of Ref.~\onlinecite{kotov98} results in $\jpl/\jp\approx 0.59$ \cite{unpub}. Keeping this uncertainty in mind, below we will quote the coupling ratio $\jpl/\jp$ relative to its clean-limit critical-point value.
To our knowledge, there are no numerical studies of the bilayer model with finite second-neighbor coupling $\jpl'$; for most of what follows we will focus on the nearest-neighbor case with $\jpl'=0$.

%%%%%%%%%%%%%%%%%%%%%%%%%%%%%%%%%%%%%%%%%%%%%%%%%%%%%%%%%%%%%%%%

\paragraph{Quenched disorder: General considerations.--}
Typical forms of quenched disorder are non-magnetic impurities and spatially inhomogeneous exchange couplings; here we will focus on the latter, i.e., bond disorder. Previous work has shown that defect bonds produce long-ranged dipolar textures in non-collinearly ordered magnets, arising from the coupling to gapless Goldstone modes \cite{dey20}. As a result, bond disorder is a relevant perturbation to such states in space dimension $d\leq 2$ \cite{cherepanov99,hasselmann04}. Therefore the 120$^\circ$ order of the triangular-lattice Heisenberg model is unstable already at infinitesimal disorder towards a spin-glass state, but with an exponentially large correlation length at weak disorder \cite{dey20}; this then also applies to $\jpl$ disorder in the present bilayer model at small $\jp$. In contrast, a single $\jp$ defect has the same symmetry as a vacancy and induces a weaker octupolar distortion \cite{wollny11}. A finite amount of $\jp$ disorder is also expected to destabilize the 120$^\circ$ order, but only via defect pairs and therefore in an even weaker fashion \cite{maryasin13}.

Thus, any finite bond disorder will turn the small-$\jp$ phase of the model \eqref{eq:Ham} into a spin glass. At very large bond disorder, this quantum spin glass will give way to a paramagnetic state dominated by dimers primarily placed on the strong bonds, i.e., a valence-bond glass \cite{kawamura14,sheng19,kimchi18,liu18}. In contrast, the large-$\jp$ dimer magnet is gapped and therefore stable against quenched disorder. This results in the phase diagram shown in Fig.~\ref{fig:phd}(b); we note that the transition between dimer and spin-glass states can be expected to display a narrow intermediate region which is gapless but shows no spin-glass order and can be characterized as a Mott glass \cite{liyao25}. Whether this Mott glass is adiabatically connected or distinct from the valence-bond glass is not known. In our numerical simulations, we will work at intermediate levels of disorder and tune $\jpl/\jp$ to access the quantum transition between the gapped dimer paramagnet and the quantum spin glass; the Mott-glass regime is invisible to our approximation.

We note that the model \eqref{eq:Ham} has been studied in Ref.~\onlinecite{hormann20} in the presence of bond disorder, using series-expansion techniques, but only in the dimer phase of small $\jpl$.

%%%%%%%%%%%%%%%%%%%%%%%%%%%%%%%%%%%%%%%%%%%%%%%%%%%%%%%%%%%%%%%%

\paragraph{Dimer parton theory.--}
We tackle the model \eqref{eq:Ham} using SU(4) bond-operator theory \cite{bop}, generalized to ordered phases \cite{bop_gen,larged1,larged2}. This can be implemented in real space for spatially inhomogeneous systems \cite{mv13}, with disorder averages taken numerically.
The formulation uses the four states of each dimer $i$ labelled $|t_k\rangle_i$,
$k=0, \ldots, 3$, where
$|t_0\rangle=
(|\!\uparrow\downarrow\rangle -|\!\downarrow\uparrow\rangle)/\sqrt{2}$,
$|t_1\rangle =
(-|\!\uparrow\uparrow\rangle+|\!\downarrow\downarrow\rangle)/\sqrt{2}$,
$|t_2\rangle =
i(|\!\uparrow\uparrow\rangle+|\!\downarrow\downarrow\rangle)/\sqrt{2}$,
$|t_3\rangle =
(|\!\uparrow\downarrow\rangle+|\!\downarrow\uparrow\rangle)/\sqrt{2}$.
While the paramagnetic dimer phase can be accessed by expanding about a product state of singlets $|t_0\rangle$, magnetic order necessitates an SU(4) rotation in the space spanned by the $|t_k\rangle$ \cite{bop_gen}, in general form \cite{mv13}
\begin{equation}
\label{trafo}
|\tilde{t}_k\rangle_i = U_{kk'}^{(i)} |t_{k'}\rangle_i,~(k,k'=0,\ldots,3).
\end{equation}
The bond-operator method starts by determining the optimal dimer product state for a given model and disorder realization, $|\tilde{\psi}_0\rangle = \prod_i |\tilde{t}_0\rangle_i$.
Then, one computes Gaussian fluctuations around this saddle-point state. To this end, one introduces bosonic excitation operators $\tilde{t}_{i\alpha}$, $\alpha=1,2,3$, defined by $|\tilde{t}_\alpha\rangle_i = \tilde{t}_{i\alpha}^\dagger |\tilde{t}_0\rangle_i$ which obey a hard-core constraint $\sum_\alpha  \tilde{t}^\dagger_{i\alpha} \tilde{t}_{i\alpha} \leq 1$ \cite{kotov98}. The Hamiltonian can be re-written in terms of the $\tilde{t}_{i\alpha}$, and the bilinear piece describes non-interacting magnetic excitation on top of $|\tilde{\psi}_0\rangle$; in a paramagnetic dimer phase these are spin-$1$ triplons. These fluctuations can be used to compute, e.g., the dynamic spin structure factor. They can also be used to compute corrections to thermodynamic observables, the leading order of which is determined by $|\tilde{\psi}_0\rangle$. Going beyond the linear approximation, this scheme can be cast into a systematic $1/d$ expansion \cite{larged1,larged2}. Technical details are given in the supplement \cite{suppl}.

%%%%%%%%%%%%%%%%%%%%%%%%%%%%%%%%%%%%%%%%%%%%%%%%%%%%%%%%%%%%%%%%

\paragraph{Phases and QPT.--}
In the following we present and discuss our numerical results. Unless noted otherwise, we utilize box-type disorder distributions for $K_{ij}$ and $J_{ij}$ with mean values $K$ and $J$ and relative widths $\lambda_K = \Delta K/K$ and $\lambda_J = \Delta J/J$, respectively. We primarily focus on the case of intralayer bond disorder, $\lambda_K\neq 0$, which -- if sufficiently strong -- produces clear signatures of spin glass-behavior even for moderate system size. We set $J=1$ and tune the dimer-to-glass transition by varying $K$; the clean-limit dimer-to-spiral transition occurs at $\Kc=1/3$.

\begin{figure}[!tb]
\includegraphics[width=\columnwidth]{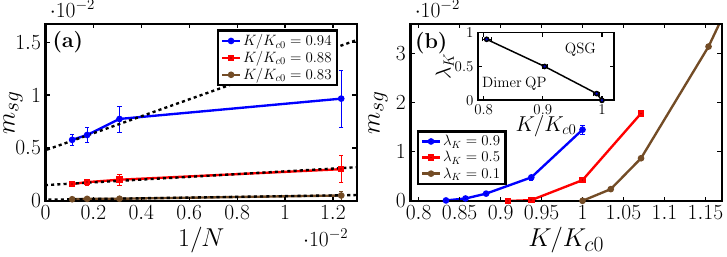}
\caption{
Spin-glass (E-A) order parameter $\msg$, obtained at the saddle-point level for intralayer box disorder.
(a) Finite-size scaling of $\msg$ for disorder strength $\lambda_K=0.9$ and different $K/\Kc$, with $N=L^2$. The dashed lines indicate a linear fit for the three largest systems.
(b) $\msg$, extrapolated to $L\to\infty$, as function of $K$ for different $\lambda_K$. The inset shows the extracted quantitative phase diagram.
The dimer--glass transition is located at $K_c/\Kc=0.991(5)$,  $K_c/\Kc=0.903(4)$ and $0.805(8)$, for $\lambda_K=0.1$, $0.5$ and $0.9$, respectively.
%\todo{reconsider values -- this is $\msg$ fit only}
We recall that $\Kc = J/3$ for the employed Gaussian approximation.
}
\label{fig:msg}
\end{figure}

We start by locating the QPT between the dimer paramagnet and the spin glass at the saddle-point level. To this end, we use $|\tilde{\psi}_0\rangle$ to compute the static observables such as local magnetizations, spin structure factor, as well as the spin-glass susceptibility and the associated Edwards-Anderson (E-A) order parameter \cite{suppl}. Sample results for the E-A order parameter of the spin glass are shown in Fig.~\ref{fig:msg}. From the finite-size scaling, we obtain the phase boundary, Fig.~\ref{fig:msg}(b). Notably, the phase transition remains well defined for any bounded disorder distribution, as the system is guaranteed to be gapped and paramagnetic for sufficiently small $K$.

The finite-size scaling of the static spin structure factor confirms the absence of long-range magnetic order in the spin-glass state at intermediate to strong disorder, with data shown in the supplement \cite{suppl}; for weak disorder the magnetic correlation length is larger than the available system sizes as discussed for the single-layer magnet \cite{dey20}.

\begin{figure}[!bt]
\includegraphics[width=\columnwidth]{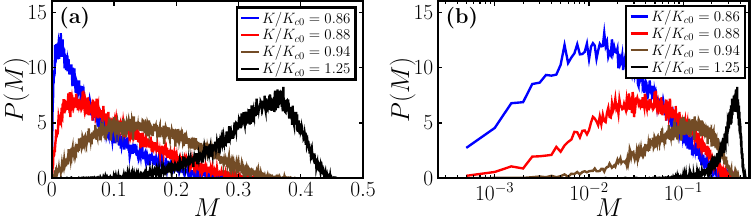}
\caption{
Histograms of the local-moment amplitude $M$, obtained at the saddle-point level for intralayer box disorder with $\lambda_K=0.9$ from 50 system realizations with $L=36$.
(a) Distribution of $M$ for different $K/\Kc$.
(b) Same distributions plotted on a logarithmic $M$ scale.
% Normalization: \int dM P(M) = 1
}
\label{fig:mdistr}
\end{figure}

The distribution of local-moment amplitudes $M_{ni}=|\langle S_{ni} \rangle|$ is shown in Fig.~\ref{fig:mdistr}. Not unexpectedly, this is narrow for parameters deep in the spin-glass phase, but becomes broad on a logarithmic scale near the QPT. Snapshots of the spatial moment distribution are shown in the supplement \cite{suppl}, indicating the formation of ordered islands forming rare events in the Griffiths sense \cite{griffiths69}. The $M$ distributions can also be used to locate the QPT, with results \cite{suppl} consistent with those from the E-A order parameter in Fig.~\ref{fig:msg}.

Together, this illustrates a strongly inhomogeneous distribution of the E-A order parameter near the transition and implies that the QPT is dominated by rare-region effects. The general arguments of Ref.~\onlinecite{tvojta06} suggest power-law quantum criticality, but it is open whether they apply to the present spin-glass case. Unfortunately, we are unable to determine the asymptotic critical behavior, due to both limitations in system size and the mean-field nature of our approximation; we therefore leave this for future work.

%%%%%%%%%%%%%%%%%%%%%%%%%%%%%%%%%%%%%%%%%%%%%%%%%%%%%%%%%%%%%%%%%%%%%%%

\begin{figure}[!b]
\includegraphics[width=\columnwidth]{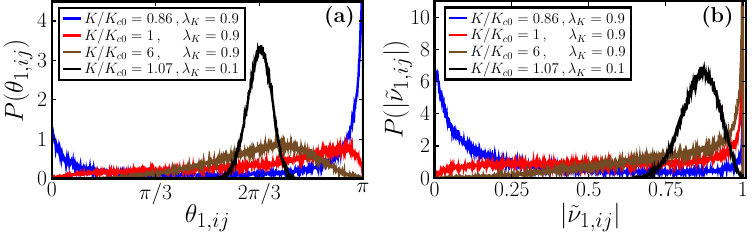}
\caption{
(a) Histograms of angles $|\theta_{1,ij}|$ between neighboring moments, obtained at the saddle-point level for intralayer box disorder from 50 system realizations with $L=30$. Data are shown for cases with both weak ($\lambda_K=0.1$) and strong disorder ($\lambda_K=0.9$).
(b) Same data, but now shown as distribution of the normalized vector chirality $\tilde\nu_{1,ij} = \nu_{1,ij}/(|\vec{S}_{1i}||\vec{S}_{1j}|)$.
}
\label{fig:thetadistr}
\end{figure}

\paragraph{Spin chiralities.--}
We employ the saddle-point state $|\tilde{\psi}_0\rangle$ for a more detailed characterization of the near-critical spin glass.
First, we consider the scalar chiralities $\kappa_{n,ijk} = (\vec{S}_{ni} \times \vec{S}_{nj})\cdot \vec{S}_{nk}$, defined on an elementary triangle $(ijk)$ in layer $n$, which quantify the degree of non-coplanarity. Distributions of $\kappa_{ijk}$, shown in the supplement, indicate that the clean-limit coplanar 120$^\circ$ state is rendered non-coplanar by bond disorder \cite{dey20}. The non-coplanarity appears stronger deep in the spin-glass phase.

\begin{figure*}[!t]
\includegraphics[width=\textwidth]{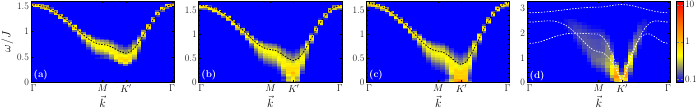}
\caption{
Dynamical spin susceptibility $\chi''(\vec{k},\w)$ for intralayer box disorder with $\lambda_K=0.9$, shown along a path in the 2D Brillouin zone with $k_z=\pi$, obtained from 25 disorder realizations with $L=30$. The discrete $\delta$ peaks have been binned with a frequency resolution of $0.03 J$ ($0.06 J$) in panels a-c (d), respectively.
(a) $K/\Kc = 0.67$,
(b) $K/\Kc = 0.79$,
(c) $K/\Kc = 0.86$,
(d) $K/\Kc = 3.0$.
Dashed lines indicate the mode dispersion in the clean limit.
}
\label{fig:spinsusc}
\end{figure*}

Second, we analyze angles between neighboring spins $\theta_{n,ij}$ and the resulting vector chiralities $\vec{\nu}_{n,ij} = \vec{S}_{ni} \times \vec{S}_{nj}$ which signal non-collinearities. Results are shown in Fig.~\ref{fig:thetadistr}. For large $K/J$ the distributions display maxima corresponding to 120$^\circ$ order which get progressively smeared with disorder. However, upon approaching the spin-glass QCP with decreasing $K$ the character of the distribution changes qualitatively, with maxima now occurring at relative angles $\theta_{ij}=0,\pi$, i.e, $\vec{\nu}_{ij}=0$.
The remarkable conclusion is that the system is more collinear and hence effectively \textit{less} frustrated close to QPT as compared to deep inside the spin-glass phase. This can be rationalized by the fact that, near the QPT, large moments are sparse due to the broad distribution of moment sizes. Therefore finding three moments of comparable size in an individual triangle -- which is the source of frustration -- is unlikely. The dominant contributions to the exchange energy instead come from configurations with one or two large moments per triangle, leading to locally collinear configurations.

%%%%%%%%%%%%%%%%%%%%%%%%%%%%%%%%%%%%%%%%%%%%%%%%%%%%%%%%%%%%%%%%%%%%%%%

\paragraph{Glassiness.--}
To probe the energy landscape of the quantum spin glass, we sample different saddle points, obtained from iterations for $|\tilde\psi_0\rangle$ with different initializations, for fixed realizations of quenched disorder. Distributions of the saddle-point energies, shown in the supplement \cite{suppl}, show the expected complex landscape with multiple inequivalent minima deep in the glass phase, but glassiness becomes weak upon approaching the QPT to the dimer phase.

As the above analysis shows, this is the result of two tendencies: Average ordered moments $M_{ni}$ are small, and non-collinearities are suppressed. Consequently, the system is only weakly frustrated, and energy barriers between the multiple stable states are low. In this sense, glassiness is ``doubly weak'' near the QPT. We believe this constitutes a general property of quantum spin-glass transitions of types (ii) and (iv) mentioned in the introduction. In contrast, glassiness is only ``singly weak'' at type-(i) transitions where moments are not small, and glassiness is strong across type-(iii) transitions which connect spin and valence-bond glasses.

%%%%%%%%%%%%%%%%%%%%%%%%%%%%%%%%%%%%%%%%%%%%%%%%%%%%%%%%%%%%%%%%%%%%%%%

\paragraph{Spin dynamics.--}
We now discuss the magnetic excitations of the system, obtained via the generalized bond-operator approach. We compute the dynamic spin susceptibility $\chi_{s,\alpha\beta}(\vec k,\w)=\llangle S^\alpha(\vec k); S^\beta(-\vec k)\rrangle_\omega$ in the single-mode approximation; in an ordered phase, this contains contributions from both transverse and longitudinal spin excitations \cite{suppl}.
Results for $\chi_s''(\vec k,\w)$ along a path in the 2D Brillouin zone (with $k_z=\pi$) are shown in Fig.~\ref{fig:spinsusc} across the phase diagram. In all cases, low-energy excitations exist primarily around $\vec k=\vec Q=(4\pi/3,0,\pi)$, the ordering wavevector of the $120^\circ$ state.
Deep in both phases, Fig.~\ref{fig:spinsusc}(a,d), the main signature of box-distributed bond disorder is a disorder-induced broadening of main modes; deep in the ordered phase, Fig.~\ref{fig:spinsusc}(d), we also observe a loss of spectral weight near the $\Gamma$ point. For bimodal disorder, with data shown in the supplement \cite{suppl}, we observe spectral weight split from -- and, depending on the type of disorder, either above or below -- the main bands.
For $K\lesssim\Kc$ box disorder fills the gap and induces a weakly ordered spin glass, Fig.~\ref{fig:spinsusc}(b,c), with the low-energy modes being particularly broad in momentum space. This indicates strong tendencies towards spatial localization, and we have confirmed this by computing the inverse participation ratio \cite{suppl}. We note that almost all intensity in $\chi_s''$ arises from transverse modes; the longitudinal (or Higgs) mode is only visible for weak disorder and near the QPT.

Due to the broken SU(2) symmetry, the spectrum is gapless anywhere in the spin-glass phase. A detailed analysis of the low-energy density of states can be found in the supplement \cite{suppl} where we also discuss the relation to the hydrodynamic theory of spin glasses \cite{saslow77}.

%%%%%%%%%%%%%%%%%%%%%%%%%%%%%%%%%%%%%%%%%%%%%%%%%%%%%%%%%%%%%%%%%%%%%%%

\paragraph{Higgs-mode dynamics.--}
Amplitude fluctuations arise primarily from local singlet formation. We therefore compute the singlet (or bond) susceptibility $\chi_b(\vec q,\w)=\llangle B(\vec q) B(-\vec q)\rrangle_\omega$ where $B_i={\vec S}_{1i}\cdot{\vec S}_{2i}$ and $\vec q$ is now a two-dimensional wavevector corresponding to the lattice of dimers. $\chi_b$ has been shown to be more sensitive to the Higgs mode of the clean-limit O(3) transition \cite{podolsky11,podolsky12,QMC_square,qin17}.

Sample results for $\chi_b''(\vec q,\w)$ are shown in Fig.~\ref{fig:bondsusc} and in the supplement \cite{suppl}.
For the clean system, an amplitude mode is clearly seen throughout the ordered phase; we note its damping due to decay into transverse spin fluctuations is neglected in the present approximation.
With intralayer disorder present, the Higgs-mode intensity in $\chi_b''(\vec q,\w)$ is drastically reduced -- an effect which does not occur for interlayer disorder. The explanation is that interlayer disorder alone leaves the $\Ztwo$ layer inversion symmetry intact, such that $\langle S_{1i}\rangle=-\langle S_{2i}\rangle$. The $B(\vec q)$ operator can therefore create a phase-coherent excitation across the lattice. For intralayer disorder, in contrast, orientational fluctuations between the layers essentially randomize the phase of the Higgs excitation; we believe this to be the primary reason of its low spectral weight.

\begin{figure}[!tb]
\includegraphics[width=\columnwidth]{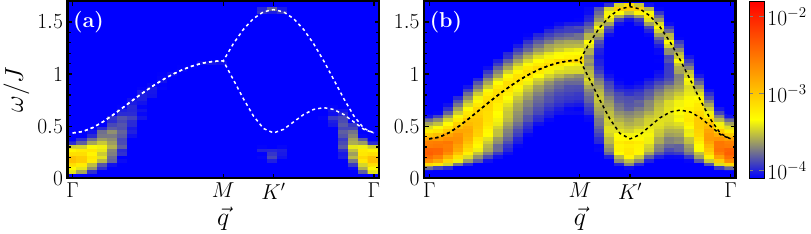}
\caption{
Dynamical bond susceptibility $\chi_b''(\vec{q},\w)$ for intralayer box disorder with $\lambda_K=0.9$, obtained from 25 disorder realizations with $L=30$.
(a) $K/\Kc = 0.81$,
(b) $K/\Kc = 0.86$,
In both cases, the overall intensity is small.
Dashed lines indicate the mode dispersion in the clean limit shifted by $\pm (4\pi/3,0)$.
}
\label{fig:bondsusc}
\end{figure}

%%%%%%%%%%%%%%%%%%%%%%%%%%%%%%%%%%%%%%%%%%%%%%%%%%%%%%%%%%%%%%%%%%%%%%%

\paragraph{Conclusions.--}
In summary, we have studied a simple case of a quantum spin-glass transition, namely the QPT between a dimer paramagnet and a spin glass, as realized in systems of magnetic dimers coupled via frustrated exchange interactions in the presence of disorder. Our model study was restricted to two space dimensions; in the three-dimensional case we expect a finite-temperature spin-glass transition \cite{maiorano18} for $K>K_{\rm c}$, with the glass temperature $T_g\to0$ as $K\to K_{\rm c}^+$.

Given that coupled-dimer magnets can be tuned by pressure between quantum paramagnetic and ordered states, the transitions advocated here should be readily accessible experimentally once bond disorder is introduced. The spin-dimer compounds Ba$_3$Mn$_2$O$_8$ \cite{stone08,samulon09},  Sr$_3$Cr$_2$O$_8$ \cite{aczel09}, as well as K$_2$Co$_2$(SeO$_3$)$_3$ \cite{fumei25} and Rb$_2$Co$_2$(SeO$_3$)$_3$ \cite{zhang26} display a layered triangular structure and can be driven from a paramagnet to canted antiferromagnet by the application of moderate magnetic fields; detailed pressure experiments have not been performed to our knowledge. Controlled chemical substitution to induce disorder should be possible; this has in fact been done for Ni(Cl$_{1-x}$Br$_x$)$_2\cdot$4SC(NH$_2$)$_2$ where bromine substitution drives a zero-field zero-pressure QPT from a paramagnet to an ordered state, with unusual transition properties reported \cite{povarov17,mannig18}. Experimentally, measurements of the NMR relaxation as well as the frequency-dependent ac susceptibility should be suitable to deduce the amount of glassiness.

%\todo{connect to various other recent papers, SYK style and otherwise}

Direct extensions of our work include (a) to analyze fluctuation corrections in detail, and (b) to include a second-neighbor coupling $K'$. With non-zero $K'$ the clean limit features a spin-liquid phase, and it will be interesting to investigate its fate in the bilayer setting with quenched disorder. Here, also the physics of valence-bond-glass (or random-singlet) states becomes pertinent \cite{kimchi18,liu18} which, however, requires an approach beyond the bond-operator technique used here. On the methodological front, we envision that the bond-operator technique can be generalized to access the real-time evolution after a quantum quench. More broadly, more investigations into quantum spin-glass criticality in realistic lattice models are called for, with one interesting question being whether local models, e.g., of Sachdev-Ye-Kitaev type, can capture the dynamics at intermediate energies or temperatures \cite{christos22,marino26}.

%%%%%%%%%%%%%%%%%%%%%%%%%%%%%%%%%%%%%%%%%%%%%%%%%%%%%%%%%%%%%%%%%%%%%%%
\acknowledgments

We thank E. C. Andrade, P. M. C\^onsoli, and S. Dey for discussions and collaborations on related work.
DGJ acknowledges support from the Department of Atomic Energy, Government of India, under Project Identification No. RTI 40007.
MV has been supported by the Deutsche Forschungsgemeinschaft through SFB 1143 (project-id 247310070) and the W\"urzburg-Dresden Cluster of Excellence on Complexity, Topology and Dynamics in Quantum Matter -- \textit{ctd.qmat} (EXC 2147, project-id 390858490).

%%%%%%%%%%%%%%%%%%%%%%%%%%%%%%%%%%%%%%%%%%%%%%%%%%%%%%%%%%%%%%%%%%%%%%%%%
%\bibliographystyle{unsrt}
\bibliography{random_tri}

@article{kotov98,
  title = {Novel Approach to Description of Spin-Liquid Phases in Low-Dimensional Quantum Antiferromagnets},
  author = {Kotov, V. N. and Sushkov, O. P. and Zheng, W. H. and Oitmaa, J.},
  journal = {Phys. Rev. Lett.},
  volume = {80},
  pages = {5790},
  year = {1998},
  publisher = {American Physical Society},
  doi = {10.1103/PhysRevLett.80.5790},
  url = {https://link.aps.org/doi/10.1103/PhysRevLett.80.5790}
}

@article{mv13,
  title = {Excitation Spectra of Disordered Dimer Magnets Near Quantum Criticality},
  author = {Vojta, Matthias},
  journal = {Phys. Rev. Lett.},
  volume = {111},
  pages = {097202},
  year = {2013},
  publisher = {American Physical Society},
  doi = {10.1103/PhysRevLett.111.097202},
  url = {https://link.aps.org/doi/10.1103/PhysRevLett.111.097202}
}

@article{dey20,
  title = {Destruction of long-range order in noncollinear two-dimensional antiferromagnets by random-bond disorder},
  author = {Dey, Santanu and Andrade, Eric C. and Vojta, Matthias},
  journal = {Phys. Rev. B},
  volume = {101},
  issue = {2},
  pages = {020411},
  numpages = {5},
  year = {2020},
  month = {Jan},
  publisher = {American Physical Society},
  doi = {10.1103/PhysRevB.101.020411},
  url = {https://link.aps.org/doi/10.1103/PhysRevB.101.020411}
}

@article{bop,
  title = {Bond-operator representation of quantum spins: Mean-field theory of frustrated quantum {Heisenberg} antiferromagnets},
  author = {Sachdev, Subir and Bhatt, R. N.},
  journal = {Phys. Rev. B},
  volume = {41},
  issue = {13},
  pages = {9323},
  year = {1990},
  doi = {10.1103/PhysRevB.41.9323},
  url = {https://link.aps.org/doi/10.1103/PhysRevB.41.9323}
}

@article{bop_gen,
	author = {Sommer, T. and Vojta, M. and Becker, K. W.},
	title = {Magnetic properties and spin waves of bilayer magnets in a uniform field},
	DOI= "10.1007/s100510170052",
	url= "https://doi.org/10.1007/s100510170052",
	journal = {Eur. Phys. J. B},
	year = 2001,
	volume = 23,
	number = 3,
	pages = "329",
	month = "",
}

@article{larged1,
  title = {Nonlinear bond-operator theory and $1/d$ expansion for coupled-dimer magnets. {I.} {P}aramagnetic phase},
  author = {Joshi, Darshan G. and Coester, Kris and Schmidt, Kai P. and Vojta, Matthias},
  journal = {Phys. Rev. B},
  volume = {91},
  issue = {9},
  pages = {094404},
  numpages = {21},
  year = {2015},
  month = {Mar},
  publisher = {American Physical Society},
  doi = {10.1103/PhysRevB.91.094404},
  url = {https://link.aps.org/doi/10.1103/PhysRevB.91.094404}
}

@article{larged2,
  title = {Nonlinear bond-operator theory and $1/d$ expansion for coupled-dimer magnets. {II.} {A}ntiferromagnetic phase and quantum phase transition},
  author = {Joshi, Darshan G. and Vojta, Matthias},
  journal = {Phys. Rev. B},
  volume = {91},
  issue = {9},
  pages = {094405},
  numpages = {19},
  year = {2015},
  month = {Mar},
  publisher = {American Physical Society},
  doi = {10.1103/PhysRevB.91.094405},
  url = {https://link.aps.org/doi/10.1103/PhysRevB.91.094405}
}

@article{wollny11,
  title = {Fractional Impurity Moments in Two-Dimensional Noncollinear Magnets},
  author = {Wollny, Alexander and Fritz, Lars and Vojta, Matthias},
  journal = {Phys. Rev. Lett.},
  volume = {107},
  issue = {13},
  pages = {137204},
  numpages = {4},
  year = {2011},
  month = {Sep},
  publisher = {American Physical Society},
  doi = {10.1103/PhysRevLett.107.137204},
  url = {https://link.aps.org/doi/10.1103/PhysRevLett.107.137204}
}

@article{maryasin13,
  title = {Triangular Antiferromagnet with Nonmagnetic Impurities},
  author = {Maryasin, V. S. and Zhitomirsky, M. E.},
  journal = {Phys. Rev. Lett.},
  volume = {111},
  issue = {24},
  pages = {247201},
  numpages = {5},
  year = {2013},
  month = {Dec},
  publisher = {American Physical Society},
  doi = {10.1103/PhysRevLett.111.247201},
  url = {https://link.aps.org/doi/10.1103/PhysRevLett.111.247201}
}

@misc{unpub,
	author = {Joshi, D. G. and Vojta, M.},
	title = {unpublished},
	year = {2026}
}

@article{QMC_square,
  title = {Dynamical structure factors and excitation modes of the bilayer {Heisenberg} model},
  author = {Loh\"ofer, M. and Coletta, T. and Joshi, D. G. and Assaad, F. F. and Vojta, M. and Wessel, S. and Mila, F.},
  journal = {Phys. Rev. B},
  volume = {92},
  issue = {24},
  pages = {245137},
  year = {2015},
  publisher = {American Physical Society},
  doi = {10.1103/PhysRevB.92.245137},
  url = {https://link.aps.org/doi/10.1103/PhysRevB.92.245137}
}

@misc{suppl,
note = {The supplementary material, which also contains Refs.~\onlinecite{matsumoto02,syrom18,puschmann20,batista25,do21}, provides details on the bond-operator technique as well as additional numerical results concerning glassiness, the static and dynamic structure factors, and the mode localization properties.},
}

@article{bernaschi24,
  title = {The quantum transition of the two-dimensional {Ising} spin glass},
  author = {Bernaschi, M. and Gonza\'lez-Adalid Pemartin, I. and Mart\'in-Mayor, V. and Parisi, G.},
  journal = {Nature (London)},
  volume = {631},
  pages = {749},
  year = {2024},
  doi = {10.1038/s41586-024-07647-y},
  url = {https://www.nature.com/articles/s41586-024-07647-y}
}

@article{rop_fru,
  title = {Frustration and quantum criticality},
  author = {Vojta, M.},
  journal = {Rep. Prog. Phys.},
  volume = {81},
  pages = {064501},
  year = {2018}
}

@article{jolicoeur90,
  title = {Ground-state properties of the {$S=1/2$} {Heisenberg} antiferromagnet on a triangular lattice},
  author = {Jolicoeur, T. and Dagotto, E. and Gagliano, E. and Bacci, S.},
  journal = {Phys. Rev. B},
  volume = {42},
  issue = {7},
  pages = {4800--4803},
  numpages = {0},
  year = {1990},
  month = {Sep},
  publisher = {American Physical Society},
  doi = {10.1103/PhysRevB.42.4800},
  url = {https://link.aps.org/doi/10.1103/PhysRevB.42.4800}
}

@article{eggert19,
  title = {Dirac Spin Liquid on the Spin-1/2 Triangular {Heisenberg} Antiferromagnet},
  author = {Hu, S. and Zhu, W. and Eggert, S. and He, Y.-C.},
  journal = {Phys. Rev. Lett.},
  volume = {123},
  pages = {207203},
  year = {2019},
  publisher = {American Physical Society},
  doi = {10.1103/PhysRevLett.123.207203},
  url = {https://link.aps.org/doi/10.1103/PhysRevLett.123.207203}
}

@article{zhuwhite15,
    title = {Spin liquid phase of the {$S=1/2$} {$J_1$-$J_2$} {Heisenberg} model on the triangular lattice},
	author = {Zhu, Z. and White, S. R.},
	journal = {Phys. Rev. B},
	volume = {92},
	pages = {041105(R)},
	year = {2015},
  doi = {10.1103/PhysRevB.92.041105},
  url = {https://link.aps.org/doi/10.1103/PhysRevB.91.041105}
}

@article{iqbal16,
    title = {Spin liquid nature in the {Heisenberg} {$J_1$-$J_2$} triangular antiferromagnet},
	author = {Iqbal, Y. and Hu, W.-J. and Thomale, R. and Poilblanc, D. and Becca, F.},
	journal = {Phys. Rev. B},
	volume = {93},
	pages = {144411},
	year = {2016},
  doi = {10.1103/PhysRevB.93.144411},
  url = {https://link.aps.org/doi/10.1103/PhysRevB.93.144411}
}

@article{ferrari19,
    title = {Dynamical structure factor of the {$J_1$-$J_2$} {Heisenberg} model on the triangular lattice: Magnons, Spinons, and Gauge fields},
	author = {Ferrari, F. and Becca, F.},
	journal = {Phys. Rev. X},
	volume = {9},
	pages = {031026},
	year = {2019},
}

@article{tang22,
    title = {Spectra of a gapped quantum spin liquid with a strong chiral excitation on the triangular lattice},
	author = {Tang, T. and Moritz, B. and Devereaux, T. P.},
	journal = {Phys. Rev. B},
	volume = {106},
	pages = {064428},
	year = {2022},
  doi = {10.1103/PhysRevB.106.064428},
  url = {https://link.aps.org/doi/10.1103/PhysRevB.106.064428}
}

@book{fischer91,
  author = {Fischer, K. H. and Hertz, J.A.},
  title = {Spin Glasses},
  publisher = {Cambridge University Press},
  year = {1991},
  address = {Cambridge}
}

@book{parisi_book,
   author = {M\'ezard, M. and Parisi, G. and Virasoro, M. A.},
   year = {1987},
   title = {Spin Glass Theory and Beyond},
   publisher = {World Scientific},
   address = {Singapore}
}

@article{villain79,
  title = {Insulating spin glasses},
  author = {Villain, J.},
  volume = {33},
  pages = {31},
  year = {1979},
  doi = {10.1007/BF01325811},
  journal = {Z. Phys. B}
}

@article{kimchi18,
 title = {Valence Bonds in Random Quantum Magnets: {T}heory and Application to {${\mathrm{YbMgGaO}}_{4}$}},
  author = {Kimchi, Itamar and Nahum, Adam and Senthil, T.},
  journal = {Phys. Rev. X},
  volume = {8},
  pages = {031028},
  year = {2018},
  publisher = {American Physical Society},
  doi = {10.1103/PhysRevX.8.031028},
  url = {https://link.aps.org/doi/10.1103/PhysRevX.8.031028}
}

@article{liu18,
 title = {Random-Singlet Phase in Disordered Two-Dimensional Quantum Magnets},
  author = {Liu, Lu and Shao, Hui and Lin, Yu-Cheng and Guo, Wenan and Sandvik, Anders W.},
  journal = {Phys. Rev. X},
  volume = {8},
  pages = {041040},
  year = {2018},
  doi = {10.1103/PhysRevX.8.041040},
  url = {https://link.aps.org/doi/10.1103/PhysRevX.8.041040}
}

@article{hormann20,
  title = {Dynamic structure factor of {Heisenberg} bilayer dimer phases in the presence of quenched disorder and frustration},
  author = {H\"ormann, M. and Schmidt, K. P.},
  journal = {Phys. Rev. B},
  volume = {102},
  pages = {094427},
  year = {2020},
  doi = {10.1103/PhysRevB.102.094427},
  url = {https://link.aps.org/doi/10.1103/PhysRevB.102.094427}
}

@article{gu00,
  title = {Order-disorder transitions in bilayer {Heisenberg} models on the triangular lattice},
  author = {Gu, Q. and Shen, J.-L.},
  journal = {Phys. Rev. B},
  volume = {62},
  pages = {3287},
  year = {2000},
  publisher = {American Physical Society},
  doi = {10.1103/PhysRevB.62.3287},
  url = {https://link.aps.org/doi/10.1103/PhysRevB.62.3287}
}

@article{singh98,
  title = {Quantum Phase Transitions in the Triangular-Lattice Bilayer {Heisenberg} Model},
  author = {Singh, R. R. P. and Elstner, N.},
  journal = {Phys. Rev. Lett.},
  volume = {81},
  pages = {4732},
  year = {1998},
  publisher = {American Physical Society},
  doi = {10.1103/PhysRevLett.81.4732},
  url = {https://link.aps.org/doi/10.1103/PhysRevLett.81.4732}
}

@article{kawamura92,
  title = {Monte {Carlo} Study of Chiral Criticality -- {XY} and {Heisenberg} Stacked-Triangular Antiferromagnets},
  author = {Kawamura, H.},
  volume = {61},
  pages = {1299},
  year = {1992},
  doi={10.7566/JPSJ.61.1299},
  journal = {J. Phys. Soc. Jpn}
}

@article{qin17,
  title = {Amplitude Mode in Three-Dimensional Dimerized Antiferromagnets},
  author = {Qin, Y. Q. and Normand, B. and Sandvik, A. W. and Meng, Z. Y.},
  volume = {118},
  pages = {147207},
  year = {2017},
  journal = {Phys. Rev. Lett},
  doi = {10.1103/PhysRevLett.118.147207},
  url = {https://link.aps.org/doi/10.1103/PhysRevLett.118.147207}
}

@article{cherepanov99,
       author = {{Cherepanov}, V. and {Korenblit}, I.~Y. and {Aharony}, A. and {Entin-Wohlman}, O.},
        title = {Suppression of antiferromagnetic correlations by quenched dipole-type impurities},
      journal = {Eur. Phys. J. B},
         year = 1999,
       volume = {8},
        pages = {511},
          doi = {10.1007/s100510050719},
       adsurl = {https://ui.adsabs.harvard.edu/abs/1999EPJB....8..511C},
}

@article{hasselmann04,
   title = {Spin-glass phase of cuprates},
  author = {Hasselmann, N. and Castro Neto, A. H. and Morais Smith, C.},
  journal = {Phys. Rev. B},
  volume = {69},
  pages = {014424},
  year = {2004},
  publisher = {American Physical Society},
  doi = {10.1103/PhysRevB.69.014424},
  url = {https://link.aps.org/doi/10.1103/PhysRevB.69.014424}
}

@article{bray80,
  title = {Replica theory of quantum spin glasses},
  author = {Bray, A. J. and Moore, M. A.},
  journal = {J. Phys. C},
  volume = {13},
  pages = {L655},
  year = {1980},
  doi = {10.1088/0022-3719/13/24/005},
  url = {https://iopscience.iop.org/article/10.1088/0022-3719/13/24/005}
}

@article{mezard84,
  title = {Nature of the Spin-Glass Phase},
  author = {Mezard, M. and Parisi, G. and Sourlas, N. and Toulouse, G. and Virasoro, M.},
  journal = {Phys. Rev. Lett.},
  volume = {52},
  pages = {1156},
  year = {1984},
  doi = {10.1103/PhysRevLett.52.1156},
  url = {https://link.aps.org/doi/10.1103/PhysRevLett.52.1156}
}

@article{huse93,
  title = {Zero-temperature critical behavior of the infinite-range quantum {Ising} spin glass},
  author = {Miller, J. and Huse, D. A.},
  journal = {Phys. Rev. Lett.},
  volume = {70},
  pages = {3147},
  year = {1993},
  publisher = {American Physical Society},
  doi = {10.1103/PhysRevLett.70.3147},
  url = {https://link.aps.org/doi/10.1103/PhysRevLett.70.3147}
}

@article{rieger94,
  title = {Zero-temperature quantum phase transition of a two-dimensional {Ising} spin glass},
  author = {Rieger, H. and Young, A. P.},
  journal = {Phys. Rev. Lett.},
  volume = {72},
  pages = {4141},
  year = {1994},
  publisher = {American Physical Society},
  doi = {10.1103/PhysRevLett.72.4141},
  url = {https://link.aps.org/doi/10.1103/PhysRevLett.72.4141}
}

@article{georges01,
  title = {Quantum fluctuations of a nearly critical {Heisenberg} spin glass},
  author = {Georges, A. and Parcollet, O. and Sachdev, S.},
  journal = {Phys. Rev. B},
  volume = {63},
  pages = {134406},
  year = {2001},
  doi = {10.1103/PhysRevB.63.134406},
  url = {https://doi.org/10.1103/PhysRevB.63.134406}
}

@article{biroli02,
  title = {Out-of-equilibrium dynamics of a quantum {Heisenberg} spin glass},
  author = {Biroli, G. and Parcollet, O.},
  journal = {Phys. Rev. B},
  volume = {65},
  pages = {094414},
  year = {2002},
  doi = {10.1103/PhysRevB.65.094414},
  url = {https://doi.org/10.1103/PhysRevB.65.094414}
}

@article{miyazaki13,
  title = {Real-space renormalization-group approach to the random transverse-field {Ising} model in finite dimensions},
  author = {Miyazaki, R. and Nishimori, H.},
  journal = {Phys. Rev. E},
  volume = {87},
  pages = {032154},
  year = {2013},
  doi = {10.1103/PhysRevE.87.032154},
  url = {https://doi.org/10.1103/PhysRevE.87.032154}
}

@article{matoz16,
  title = {Unconventional critical activated scaling of two-dimensional quantum spin glasses},
  author = {Matoz-Fernandez, D. A. and Roma, F.},
  journal = {Phys. Rev. B},
  volume = {94},
  pages = {024201},
  year = {2016},
  doi = {10.1103/PhysRevB.94.024201},
  url = {https://doi.org/10.1103/PhysRevB.94.024201}
}

@article{arrachea02,
  title = {Infinite-range quantum random {Heisenberg} magnet},
  author = {Arrechea, L. and Rozenberg, M. J.},
  journal = {Phys. Rev. B},
  volume = {65},
  pages = {224430},
  year = {2002},
  doi = {10.1103/PhysRevB.65.224430},
  url = {https://doi.org/10.1103/PhysRevB.65.224430}
}

@article{saslow77,
  title = {Hydrodynamic theory of spin waves in spin glasses and other systems with noncollinear spin orientations},
  author = {Halperin, B. I. and Saslow, W. M.},
  journal = {Phys. Rev. B},
  volume = {16},
  pages = {2154},
  year = {1977},
  doi = {10.1103/PhysRevB.16.2154},
  url = {https://doi.org/10.1103/PhysRevB.16.2154}
}

@article{baity13,
  title = {Critical parameters of the three-dimensional {Ising} spin glass},
  author = {Baity-Jesi, M. and others},
  journal = {Phys. Rev. B},
  volume = {88},
  pages = {224416},
  year = {2013},
  doi = {10.1103/PhysRevB.88.224416},
  url = {https://doi.org/10.1103/PhysRevB.88.224416}
}

@article{christos22,
  title = {Spin liquid to spin glass crossover in the random quantum {Heisenberg} magnet},
  author = {Christos, M. and Haehl, F. M. and Sachdev, S.},
  journal = {Phys. Rev. B},
  volume = {105},
  pages = {085120},
  year = {2022},
  publisher = {American Physical Society},
  doi = {10.1103/PhysRevB.105.085120},
  url = {https://doi.org/10.1103/PhysRevB.105.085120}
}

@article{franz94,
  title = {Interfaces and lower critical dimension in a spin glass model},
  author = {Franz, S. and Parisi, G. and Virasoro, M.-A.},
  journal = {J. Phys.},
  volume = {4},
  pages = {1657},
  year = {1994}
}

@article{maiorano18,
  title = {Support for the value 5/2 for the spin-glass lower critical dimension at zero magnetic field},
  author = {Maiorano, A. and Parisi, G.},
  journal = {Proc. Natl. Acad. Sci. USA},
  volume = {115},
  pages = {5129},
  year = {2018},
  doi = {10.1073/pnas.1720832115},
  url = {https://doi.org/10.1073/pnas.1720832115}
}

@article{angelini22,
    title = {Unexpected Upper Critical Dimension for Spin Glass Models in a Field Predicted by the Loop Expansion around the Bethe Solution at Zero Temperature},
    author = {Angelini, M. C. and Lucibello, C. and Parisi, G. and Perrupato, G. and Ricci-Tersenghi, F. and Rizzo, T.},
    year = {2022},
    journal = {Phys. Rev. Lett.},
    volume = {128},
    pages = {075702},
    doi = {10.1103/PhysRevLett.128.075702},
    url = {https://doi.org/10.1103/PhysRevLett.128.075702}
}

@article{sherrington75,
    title = {Solvable Model of a Spin Glass},
    author = {Sherrington, D. and Kirkpatrick, S.},
    year = {1975},
    journal = {Phys. Rev. Lett.},
    volume = {35},
    pages = {1792},
    doi = {10.1103/PhysRevLett.35.1792},
    url = {https://doi.org/10.1103/PhysRevLett.35.1792}
}

@article{edwards75,
    title = {Theory of spin glasses},
    author = {Edwards, S. F. and Anderson, P. W.},
    year = {1975},
    journal = {J. Phys. F: Met. Phys.},
    volume = {5},
    pages = {965},
    doi = {10.1088/0305-4608/5/5/017},
    url = {https://doi.org/10.1088/0305-4608/5/5/017}
}

@article{georges00,
    title = {Mean Field Theory of a Quantum {Heisenberg} Spin Glass},
    author = {Georges, A. and Percollet, O. and Sachdev, S.},
    year = {2000},
    journal = {Phys. Rev. Lett.},
    volume = {85},
    pages = {840},
    doi = {10.1103/PhysRevLett.85.840},
    url = {https://doi.org/10.1103/PhysRevLett.85.840}
}

@article{shackleton21,
    title = {Quantum Phase Transition at Nonzero Doping in a Random {$t$-$J$} Model},
    author = {Shackleton, H. and Wietek, A. and Georges, A. and Sachdev, S.},
    year = {2021},
    journal = {Phys. Rev. Lett.},
    volume = {126},
    pages = {136602},
    doi = {10.1103/PhysRevLett.126.136602},
    url = {https://doi.org/10.1103/PhysRevLett.126.136602}
}

@article{kavokine24,
    title = {Exact Numerical Solution of the Fully Connected Classical and Quantum {Heisenberg} Spin Glass},
    author = {Kavokine, N. and M\"uller, M. and Georges, A. and Percollet, O.},
    year = {2024},
    journal = {Phys. Rev. Lett.},
    volume = {133},
    pages = {016501},
    doi = {10.1103/PhysRevLett.133.016501},
    url = {https://doi.org/10.1103/PhysRevLett.133.016501}
}

@article{chowdhury_rmp,
    title = {{Sachdev-Ye-Kitaev} models and beyond: Window into non-{F}ermi liquids},
    author = {Chowdhury, D. and Georges, A. and Percollet, O. and Sachdev, S.},
    year = {2022},
    journal = {Rev. Mod. Phys.},
    volume = {94},
    pages = {035004},
    doi = {10.1103/RevModPhys.94.035004},
    url = {https://doi.org/10.1103/RevModPhys.94.035004}
}

@misc{bracci26,
author = {Bracci-Testasecca, G. and Niedda, J. and Coraggio, A. and Moessner, R. and Scardicchio, A.},
title = {Semiclassical picture of the {Heisenberg} spin glass in two dimensions: from weak localization to hydrodynamics},
note = {{arXiv:2603.22077}},
year = {2026},
url = {https://doi.org/10.48550/arXiv.2603.22077}
}

@misc{viteritti25,
author = {Viteritti, L. L. and Rende, R. and Bracci-Testasecca, G. and Niedda, J. and Moessner, R. and Carleo, G. and Scardicchio, A.},
title = {Quantum Spin Glass in the Two-Dimensional Disordered {Heisenberg} Model via Foundation Neural-Network Quantum States},
note = {{arXiv:2507.05073}},
year = {2025},
url = {https://doi.org/10.48550/arXiv.2507.05073}
}

@misc{marino26,
title = {Crossover to {Sachdev-Ye-Kitaev} criticality in an infinite-range quantum {Heisenberg} spin glass},
author = {Hosseinabadi, H. and Sachdev, S. and Marino, J.},
note = {{arXiv:2603.11263}},
year = {2026},
url = {https://doi.org/10.48550/arXiv.2603.11263}
}

@misc{fumei25,
author = {Fu, Y. and Ge, H. and Chen, J. and Xiao, J. and Tan, Y. and Wang, L. and Wang, J. and Dong, C. and Qu, Z. and He, M. and Xi, C. and Ling, L. and Xi, B. and Mei, J.-W.},
title = {{Berezinskii-Kosterlitz-Thouless} region and magnetization plateaus in easy-axis triangular weak-dimer antiferromagnet {K$_2$Co$_2$(SeO$_3$)$_3$}},
note = {{arXiv:2501.09619}},
year = {2025},
url = {https://doi.org/10.48550/arXiv.2501.09619}
}

@misc{zhang26,
author = {Zhang, S. and Silva Freitas, G. and Zapf, V. S. and Lee, M. and Choi, W. and Lin, S.-Z. and Chen, T. and Broholm, C. and Xu, X. and Cava, R. J. and Choi, S.},
title = {Magnetization Plateaus in the Spin-Orbit Coupled Bilayer Triangular Lattice Antiferromagnet {Rb$_2$Co$_2$(SeO$_3$)$_3$}},
note = {{arXiv:2601.19882}},
year = {2026},
url = {https://doi.org/10.48550/arXiv.2601.19882}
}

@misc{liyao25,
author = {Li, K. and Wu, H.-Q. and Yao, D.-X.},
title = {Mott Glass and Criticality in a {$S=1/2$} Bilayer {Heisenberg} Model with Interlayer Bond Dilution},
note = {{arXiv:2509.03604}},
year = {2025},
url = {https://doi.org/10.48550/arXiv.2509.03604}
}

@article{povarov17,
  title = {Quantum criticality in a three-dimensional spin system at zero field and pressure},
  author = {Povarov,  K. Yu. and Mannig, A. and Perren, G. and M\"oller, J. S. and Wulf, E. and Ollivier, J. and Zheludev, A.},
  journal = {Phys. Rev. B},
  volume = {96},
  pages = {140414},
  year = {2017},
  doi = {10.1103/PhysRevB.96.140414},
  url = {https://link.aps.org/doi/10.1103/PhysRevB.96.140414}
}

@article{mannig18,
  title = {Spin waves near the edge of halogen substitution induced magnetic order in {Ni(Cl$_{1-x}$Br$_x$)$_2\cdot$4SC(NH$_2$)$_2$}},
  author = {Mannig, A. and Povarov,  K. Yu. and Ollivier, J. and Zheludev, A.},
  journal = {Phys. Rev. B},
  volume = {98},
  pages = {214419},
  year = {2018},
  doi = {10.1103/PhysRevB.98.214419},
  url = {https://link.aps.org/doi/10.1103/PhysRevB.98.214419}
}

@article{stone08,
    title = {Singlet-Triplet Dispersion Reveals Additional Frustration in the Triangular-Lattice Dimer Compound {Ba$_3$Mn$_2$O$_8$}},
    author = {Stone, M. B. and Lumsden, M. D. and Chang, S. and Samulon, E. C. and Batista, C. D. and Fisher, I. R.},
    journal = {Phys. Rev. Lett.},
    volume = {100},
    pages = {237201},
    year = {2008},
    doi = {10.1103/PhysRevLett.100.237201},
    url = {https://doi.org/10.1103/PhysRevLett.100.237201}
}

@article{samulon09,
    title = {Asymmetric Quintuplet Condensation in the Frustrated $S=1$ Spin Dimer Compound {Ba$_3$Mn$_2$O$_8$}},
    author = {Samulon, E. C. and Kohama, Y. and McDonald, R. D. and Shapiro, M. C. and Al-Hassanieh, K. A. and Batista, C. D. and Jaime, M. and Fisher, I. R.},
    year = {2009},
    journal = {Phys. Rev. Lett.},
    volume = {103},
    pages = {047202},
    doi = {10.1103/PhysRevLett.103.047202},
    url = {https://doi.org/10.1103/PhysRevLett.103.047202}
}

@article{aczel09,
    title = {Field-induced {Bose-Einstein} Condensation of triplons up to {8\,K} in {Sr$_3$Cr$_2$O$_8$}},
    author = {Aczel, A. A. and Kohama, Y. and Marcenat, C. and Weickert, F. and Ayala-Valenzuela, O. E. and Jaime, M. and McDonald, R. D. and Selesnic, S. D. and Dabkowska, H. A. and Luke, G. M.},
    year = {2009},
    journal = {Phys. Rev. Lett.},
    volume = {103},
    pages = {207203},
    doi = {10.1103/PhysRevLett.103.207203},
    url = {https://doi.org/10.1103/PhysRevLett.103.207203}
}

@article{kawamura14,
    title = {Quantum spin-liquid behavior in the spin-1/2 random {Heisenberg} antiferromagnet on the triangular lattice},
    author = {Watanabe, K. and Kawamura, H. and Nakano, H. and Sakai, T.},
    journal = {J. Phys. Soc. Jpn.},
    volume = {83},
    pages = {034714},
    year = {2014},
}

@article{sheng19,
    title = {Randomness-induced spin-liquid-like phase in the spin-1/2 {$J_1$-$J_2$} triangular {Heisenberg} model},
    author = {Wu, H.-Q. and Gong, S.-S. and Sheng, D. N},
    journal = {Phys. Rev. B},
    volume = {99},
    pages = {085141},
    year = {2019},
    doi = {10.1103/PhysRevB.99.085141},
    url = {https://link.aps.org/doi/10.1103/PhysRevB.99.085141}
}

@article{podolsky11,
  title = {Visibility of the amplitude ({Higgs}) mode in condensed matter},
  author = {Podolsky, Daniel and Auerbach, Assa and Arovas, Daniel P.},
  journal = {Phys. Rev. B},
  volume = {84},
  issue = {17},
  pages = {174522},
  numpages = {17},
  year = {2011},
  month = {Nov},
  publisher = {American Physical Society},
  doi = {10.1103/PhysRevB.84.174522},
  url = {https://link.aps.org/doi/10.1103/PhysRevB.84.174522}
}

@article{podolsky12,
  title = {Spectral functions of the {Higgs} mode near two-dimensional quantum critical points},
  author = {Podolsky, Daniel and Sachdev, Subir},
  journal = {Phys. Rev. B},
  volume = {86},
  issue = {5},
  pages = {054508},
  numpages = {14},
  year = {2012},
  month = {Aug},
  publisher = {American Physical Society},
  doi = {10.1103/PhysRevB.86.054508},
  url = {https://link.aps.org/doi/10.1103/PhysRevB.86.054508}
}

@article{griffiths69,
  title = {Nonanalytic Behavior Above the Critical Point in a Random {Ising} Ferromagnet},
  author = {Griffiths, Robert B.},
  journal = {Phys. Rev. Lett.},
  volume = {23},
  pages = {17},
  year = {1969},
  publisher = {American Physical Society},
  doi = {10.1103/PhysRevLett.23.17},
  url = {https://link.aps.org/doi/10.1103/PhysRevLett.23.17}
}

@article{tvojta06,
    title = {Rare region effects at classical, quantum and nonequilibrium phase transitions},
    author = {Vojta, T.},
    journal = {J. Phys. A: Math. Gen.},
    volume = {39},
    pages = {R143},
    year = {2006},
    doi = {10.1088/0305-4470/39/22/R01},
    url = {https://doi.org/10.1088/0305-4470/39/22/R01}
}

\newpage


\section*{Supplementary material}

# Supplemental material:
# Quantum spin-glass criticality in disordered frustrated dimer magnets

Darshan G. Joshi[1] and Matthias Vojta[2]

[1] *Tata Institute of Fundamental Research, Hyderabad 500046, India*
[2] *Institut für Theoretische Physik and Würzburg-Dresden Cluster of Excellence ctd.qmat, Technische Universität Dresden, 01062 Dresden, Germany*



## I. REAL-SPACE BOND PARTON THEORY

The dimer parton approach represents an expansion about dimer product states, using the SU(4) basis states $|t_{0\ldots3}\rangle_i$ per dimer $i$. This method, also dubbed bond-operator approach, has been originally developed for paramagnetic states [1] and later generalized to describe symmetry-broken states and associated quantum phase transitions [2, 3]. With quenched disorder present, states are in general inhomogeneous, and the parton theory needs to be implemented in real space on finite lattices with $N = L^d$ sites, such that disorder averages are taken numerically [4]. A closely similar approach has been used to access the Higgs-mode dynamics near the superfluid-Bose glass quantum phase transition [5].

Here we generalize the method, developed in Ref. 4 for collinear states in the presence of quenched disorder, to the non-collinear (as well as non-coplanar) case. Below we give a brief summary, highlighting the differences to Ref. 4; for a detailed exposure we refer the reader to Ref. 4.

### A. Dimer product state

The first step amounts to determining $|\tilde{\psi}_0\rangle = \prod_i |\tilde{t}_0\rangle_i$ as the best product-state approximation to the ground state of $\mathcal{H}$ for a given realization of quenched disorder. This then defines the basis rotation in the four-dimensional Hilbert space of each dimer, Eq. (2) in the main text, up to an SU(3) rotation within the excited states $|\tilde{t}_{1,2,3}\rangle$. There are multiple ways to determine the product state $|\tilde{\psi}_0\rangle$. We employ a dimer mean-field theory, with the single-dimer mean-field Hamiltonian

$$\mathcal{H}^i_{MF} = J_i \vec{S}_{1i}\cdot\vec{S}_{2i} + \sum_{jm} K_{m,ij}\vec{S}_{mi}\cdot\langle\vec{S}_{mj}\rangle\,, \tag{S1}$$

where the spin expectation value is taken with respect to the state $|\tilde{t}_0\rangle_i$. This Hamiltonian can be solved exactly. Starting from an initial guess for the spin expectation values $\langle\vec{S}_{mi}\rangle$, we iterate the $N$ coupled mean-field problems until convergence to find the state $|\tilde{t}_0\rangle_i$ as the ground state of $\mathcal{H}^i_{MF}$ for each dimer site. It is straightforward to show that this state corresponds to a variational local minimum of the energy [4], also dubbed saddle point in the following.

For non-collinear order, in general all four states $|t_{0\ldots4}\rangle$ mix; this is different from the collinear case of Ref. 4 where $|\tilde{t}_0\rangle$ can be chosen as a linear combination of $|t_0\rangle$ and $|t_3\rangle$. Although convergence of the mean-field iteration can be easily achieved in most cases, the present combination of frustration and quenched disorder leads to the existence of many saddle points which are inequivalent, i.e., not related by global SU(2) spin rotations. We therefore use a large number of initial configurations (typically $10^4$) for each realization of disorder and record their converged states and energies. The lowest-energy state is then used for further computations; the histogram of all converged energies is used to study properties of the energy landscape, i.e., its glassiness.

### B. Gaussian fluctuations

In a second step, the spin Hamiltonian $\mathcal{H}$ is expressed via bosonic operators. To this end, the spin operators are represented in terms of transition operators between the states $|t_k\rangle_i$ of a dimer,

$$S^\alpha_{mi} = \sum_{kk'} s^{\alpha m}_{kk'}|t_k\rangle_i\,{}_i\langle t_{k'}| \tag{S2}$$

with $4\times4$ matrices $s^{\alpha m}$ for the spin components $S^\alpha$ ($\alpha = x, y, z \equiv 1, 2, 3$) of the $m = 1, 2$ spins [1]:

$$s^{x1,2} = \frac{1}{2}\begin{pmatrix} 0 & \pm1 & 0 & 0 \\ \pm1 & 0 & 0 & 0 \\ 0 & 0 & 0 & -i \\ 0 & 0 & i & 0 \end{pmatrix},$$

$$s^{y1,2} = \frac{1}{2}\begin{pmatrix} 0 & 0 & \pm1 & 0 \\ 0 & 0 & 0 & i \\ \pm1 & 0 & 0 & 0 \\ 0 & -i & 0 & 0 \end{pmatrix},$$

$$s^{z1,2} = \frac{1}{2}\begin{pmatrix} 0 & 0 & 0 & \pm1 \\ 0 & 0 & -i & 0 \\ 0 & i & 0 & 0 \\ \pm1 & 0 & 0 & 0 \end{pmatrix}. \tag{S3}$$

Using the basis rotation (2), Eq. (S2) turns into

$$S^\alpha_{im} = \sum_{kk'} \tilde{s}^{\alpha m}_{i,kk'}|\tilde{t}_k\rangle_i\,{}_i\langle\tilde{t}_{k'}|\,, \tag{S4}$$

with site-dependent spin matrices

$$\tilde{s}^{\alpha m}_{i,kk'} = \sum_{ll'}(U^\dagger)^{(i)}_{lk}s^{\alpha m}_{ll'}U^{(i)}_{k'l'}. \tag{S5}$$

The full Hamiltonian can be represented exactly in terms of transition operators as

$$\mathcal{H} = \sum_{ikk'} X_i^{kk'} |\tilde{t}_k\rangle_i \, {}_i\langle \tilde{t}_{k'}| + \sum_{ii'kk'll'} Y_{ii'}^{kk'll'} |\tilde{t}_k\rangle_i |\tilde{t}_l\rangle_{i'} \, {}_i\langle \tilde{t}_{k'}| \, {}_{i'}\langle \tilde{t}_{l'}|. \tag{S6}$$

Its coefficients $X_i^{kk'}$ and $Y_{ii'}^{kk'll'}$ are proportional to the original couplings $J_i$ and $K_{m,ii'}$, respectively. Given the lengthy expressions (not given here), it is most efficient to generate them numerically for a finite-size system with a given realization of disorder and a given reference product state, using the equations (1), (S4), and (S5).

To make progress, this Hamiltonian is now rewritten in terms of $\tilde{t}_{i\alpha}$ ($\alpha = 1, 2, 3$) bosonic excitation operators which are defined to act on the background state $\prod_i |\tilde{t}_0\rangle_i$, i.e., $\tilde{t}_{i\alpha}^\dagger |\tilde{t}_0\rangle_i = |\tilde{t}_\alpha\rangle_i$. In the paramagnetic case, the $\tilde{t}_{i\alpha}$ reduce to triplon operators creating spin-1 excitations in a background of singlet dimers. Using the decomposition $\mathcal{H} = \mathcal{H}_0 + \mathcal{H}_1 + \mathcal{H}_2 + \mathcal{H}_3 + \mathcal{H}_4$ we have

$$\begin{aligned}
\mathcal{H}_0 &= \sum_i X_i^{00} + \sum_{ii'} Y_{ii'}^{0000}, \\
\mathcal{H}_1 &= \sum_{i\alpha} \left( X_i^{0\alpha} \tilde{t}_{i\alpha} + X_i^{\alpha 0} \tilde{t}_{i\alpha}^\dagger \right) + \sum_{ii'\alpha} \left( Y_{ii'}^{0\alpha 00} \tilde{t}_{i\alpha} + Y_{ii'}^{000\alpha} \tilde{t}_{i'\alpha} + Y_{ii'}^{\alpha 000} \tilde{t}_{i\alpha}^\dagger + Y_{ii'}^{00\alpha 0} \tilde{t}_{i'\alpha}^\dagger \right), \\
\mathcal{H}_2 &= \sum_{i\alpha\beta} (X_i^{\alpha\beta} - X_i^{00}\delta_{\alpha\beta}) \tilde{t}_{i\alpha}^\dagger \tilde{t}_{i\beta} + \sum_{ii'\alpha\beta} \left[ (Y_{ii'}^{\alpha\beta 00} - Y_{ii'}^{0000}\delta_{\alpha\beta}) \tilde{t}_{i\alpha}^\dagger \tilde{t}_{i\beta} (Y_{ii'}^{00\alpha\beta} - Y_{ii'}^{0000}\delta_{\alpha\beta}) \tilde{t}_{i'\alpha}^\dagger \tilde{t}_{i'\beta} \right] \\
&\quad + \sum_{ii'\alpha\beta} \left( Y_{ii'}^{\alpha 00\beta} \tilde{t}_{i\alpha}^\dagger \tilde{t}_{i'\beta} + Y_{ii'}^{0\beta\alpha 0} \tilde{t}_{i'\alpha}^\dagger \tilde{t}_{i\beta} + Y_{ii'}^{\alpha 0\beta 0} \tilde{t}_{i\alpha}^\dagger \tilde{t}_{i'\beta}^\dagger + Y_{ii'}^{0\alpha 0\beta} \tilde{t}_{i\alpha} \tilde{t}_{i'\beta} \right)
\end{aligned} \tag{S7}$$

where $\alpha, \beta = 1, 2, 3$. The Gaussian fluctuations of interest are contained in $\mathcal{H}_2$, which describes $3N$ modes where $N$ is the number of dimers, and the factor of three arises from the spin polarizations $\alpha = x, y, z$. $\mathcal{H}_1$ vanishes identically for the optimum product state; $\mathcal{H}_{3,4}$ correspond to mode–mode coupling terms and are neglected in the treatment.

We generate a matrix form of $\mathcal{H}_2$ numerically from the coefficients $X_i^{kk'}$ and $Y_{ii'}^{kk'll'}$ via Eq. (S7). This non-Hermitian matrix problem is then diagonalized by a bosonic Bogoliubov transformation [6] using standard LAPACK routines. In the paramagnetic phase, where $U$ is a unit matrix, $\mathcal{H}_2$ decomposes into three equivalent blocks, describing the triply degenerate spin-1 (or triplon) excitations. For collinear order, the modes describe decoupled longitudinal and transverse excitations. In contrast, $\mathcal{H}_2$ is no longer block-diagonal for non-collinear order, such that the full $6N \times 6N$ matrix needs to be diagonalized. The $6N$ eigenvalues come in pairs with opposite sign; in the ordered phases we always find three pairs of eigenvalues with magnitude smaller than $10^{-4}J$, corresponding to the expected zero-energy (Goldstone) modes.

The positive-energy eigenvalues $\omega_n$ and their eigenvectors $(u_n, v_n)$ represent the excitation modes $\tau_n$ of $\mathcal{H}$, being linear combinations of creation and annihilation $\tilde{t}$ operators:

$$\tau_n = \sum_{i\alpha} \left( u_{ni\alpha} \tilde{t}_{i\alpha} + v_{ni\alpha} \tilde{t}_{i\alpha}^\dagger \right). \tag{S8}$$

The $\omega_n$ and $(u_n, v_n)$ encode the Gaussian fluctuations around the saddle point described by $|\tilde{\psi}_0\rangle$ and can be utilized to extract physical observables.

### C. Observables

Using the dimer parton theory, we can compute different observables that we use as our diagnostic tools. The static spin structure factor is defined as usual,

$$S(\vec{k}) = \langle S(\vec{k}) S(-\vec{k}) \rangle = \frac{1}{2N} \left[ \sum_{ii'mm'} \langle \vec{S}_{mi} \cdot \vec{S}_{m'i'} \rangle e^{-i\vec{k}\cdot\vec{r}_{mi,m'i'}} \right]_{av}, \tag{S9}$$

with $2N$ the number of lattice sites, $\vec{r}_{mi,m'i'}$ the distance between the two spin sites, and $[...]_{av}$ denoting disorder average. Note that $\vec{k}$ is a three-component momentum vector, with $k_z = 0, \pi$ for the present bilayer geometry. In analogy, we can define a singlet structure factor

$$B(\vec{q}) = \frac{1}{N} \left[ \sum_{ij} \langle (\vec{S}_{1i} \cdot \vec{S}_{2i})(\vec{S}_{1j} \cdot \vec{S}_{2j}) \rangle e^{-i\vec{q}\cdot\vec{r}_{ij}} \right]_{av} \tag{S10}$$

where $\vec{r}_{ij}$ is now the distance between two dimer sites and $\vec{q}$ a two-dimensional momentum vector. Order parameters can be defined from the structure factor peaks, e.g., $m_S^2 = S(\vec{Q})/(2N)$ with $\vec{Q} = \{4\pi/3, 0\}$ for the 120° Néel phase. We also define a spin-glass susceptibility, here for

the staggered magnetization components of the dimers,

$$\chi_{sg}(\vec{q}) = N \sum_{\alpha,\beta=x,y,z} \left[\langle |q^{\alpha,\beta}(\vec{q})|^2\rangle\right]_{av} , \tag{S11}$$

$$q^{\alpha,\beta}(\vec{q}) = \frac{1}{N}\sum_i \langle (S_{1i}^{\alpha,(1)} - S_{2i}^{\alpha,(1)})(S_{1i}^{\beta,(2)} - S_{2i}^{\beta,(2)})\rangle \times e^{-i\vec{q}\cdot\vec{r}_i}$$

where the superscripts (1), (2) denote two different replicas of the disorder. Now, the Edwards-Anderson (E-A) order parameter of the spin glass is given by $m_{sg}^2 = \xi_{sg}(0)$.

To evaluate the structure factors, one uses the spin representation (S4) which can be cast into

$$S_{mi}^{\alpha} = \tilde{s}_{i,00}^{\alpha m} P_i + \sum_{\mu=1}^{3}\left[\tilde{s}_{i,\mu 0}^{\alpha m}\tilde{t}_{i\mu}^{\dagger} P_i + h.c.\right] + \sum_{\mu,\nu=1}^{3} \tilde{s}_{i,\mu\nu}^{\alpha m}\tilde{t}_{i\mu}^{\dagger}\tilde{t}_{i\nu} , \tag{S12}$$

where $P_i = 1 - \sum_{\gamma=1,2,3}\tilde{t}_{i\gamma}^{\dagger}\tilde{t}_{i\gamma}$. Expectation values then receive contributions from the product state $|\tilde{\psi}_0\rangle$ as well as from fluctuation corrections, the latter involving expectation values of the form $\langle \tilde{t}^{\dagger}\tilde{t}\rangle$. In most of this paper, we restrict ourselves to the leading-order contributions.

Dynamic versions of these structure factors can be defined analogously. They can be computed in a single-mode approximation from the eigenvalues and eigenvectors of the $3N \times 3N$ Bogoliubov problem, Eq. (S8), for details see Ref. 4.

## II. CLEAN TRIANGULAR BILAYER MAGNET

Given that the paper is about describing the bond-disordered triangular-lattice bilayer Heisenberg magnet using bond parton theory, we quickly summarize the corresponding procedure in the clean-limit case.

### A. Bond operators in harmonic approximation

The linearized bond-operator theory for the paramagnetic dimer phase [2, 3, 7] delivers a triplon dispersion of the form

$$\omega_{\vec{q}} = \sqrt{A_{\vec{q}}^2 - B_{\vec{q}}^2}, \quad B_{\vec{q}} = 2K\gamma_{\vec{q}}, \; A_{\vec{q}} = J + B_{\vec{q}} \tag{S13}$$

with the lattice form factor

$$\gamma_{\vec{q}} = \cos q_x + \cos\frac{q_x + \sqrt{3}q_y}{2} + \cos\frac{q_x - \sqrt{3}q_y}{2}.$$

The dispersion minimum is located at $\vec{Q} = (4\pi/3, 0)$, with a gap given by

$$\Delta = \sqrt{J - 3K}. \tag{S14}$$

This yields the location of the transition into the 120° ordered phase as $K_c/J = 1/3$.

To describe the non-collinear 120° phase, here with spins in the $xy$ plane, requires to introduce a condensate of $|t_1\rangle$ and $|t_2\rangle$ as follows

$$\begin{aligned}
|\tilde{t}_0\rangle_i &= \frac{|t_0\rangle_i + \lambda\big(c_i|t_2\rangle_i + s_i|t_1\rangle_i\big)}{\sqrt{1+\lambda^2}} , \\
|\tilde{t}_1\rangle_i &= \frac{-\lambda|t_0\rangle_i + \big(c_i|t_2\rangle_i + s_i|t_1\rangle_i\big)}{\sqrt{1+\lambda^2}} , \\
|\tilde{t}_2\rangle_i &= s_i|t_2\rangle_i - c_i|t_1\rangle_i , \\
|\tilde{t}_3\rangle_i &= |t_3\rangle_i ,
\end{aligned} \tag{S15}$$

where $c_i = \cos\vec{Q}\cdot\vec{r}_i$, $s_i = \sin\vec{Q}\cdot\vec{r}_i$, and $\lambda$ is the condensate parameter

$$\lambda = \sqrt{\frac{3K - J}{3K + J}} . \tag{S16}$$

To obtain the dispersion of the generalized triplon modes requires the diagonalization of a $4 \times 4$ matrix (as well as of a $2 \times 2$ matrix); the former cannot be done analytically in general. We note that the structure (S15) implies that the transverse mode for out-of-plane fluctuations is decoupled, while the transverse mode for in-plane fluctuations and the longitudinal mode are coupled. The mode dispersions of the clean system are given in several figures as dashed lines for reference.

### B. Beyond the harmonic approximation

The bond-operator theory can be developed beyond the independent-boson approximation, and numerous ideas have been put forward. The hardcore repulsion of the triplons can be implemented diagrammatically using Brueckner theory which, however, only works in the dimer phase [7]. Alternatively, one can generate series expansions somewhat similar to spin-wave theory which can be cast either into an expansion in $1/d$ with $d$ the number of space dimensions [8, 9] or an expansion in $1/n$ where $n$ is the triplon occupation number [10]; these approaches work in both the paramagnetic and antiferromagnetic phases and thus capture the quantum phase transition, albeit asymptotically with mean-field exponents. Very recently, a Schwinger-boson large-$N$ expansion has been proposed which operates in the SU(4) space of dimers [11].

With quenched disorder included, none of these methods has been pursued. We leave this formidable task for future work.

## III. LOCATING THE QUANTUM PHASE TRANSITION

At the level of the saddle-point description, the transition between the quantum spin-glass and dimer states

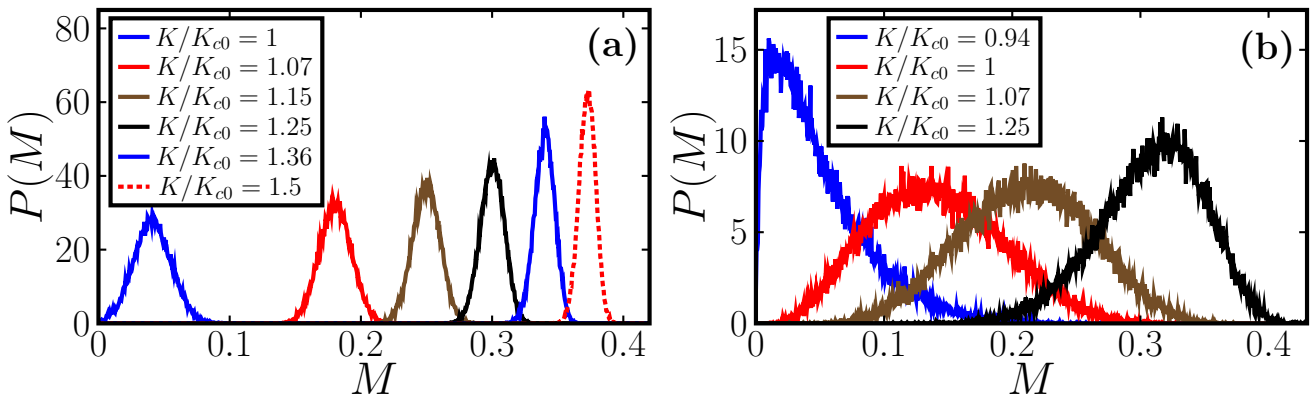

FIG. S1. Histograms of the local magnetization $M$ as in Fig. 3 of the main paper, but here for intralayer box disorder with (a) $\lambda_K = 0.1$ and (b) $\lambda_K = 0.5$, each obtained from 50 system realizations with $L = 30$.

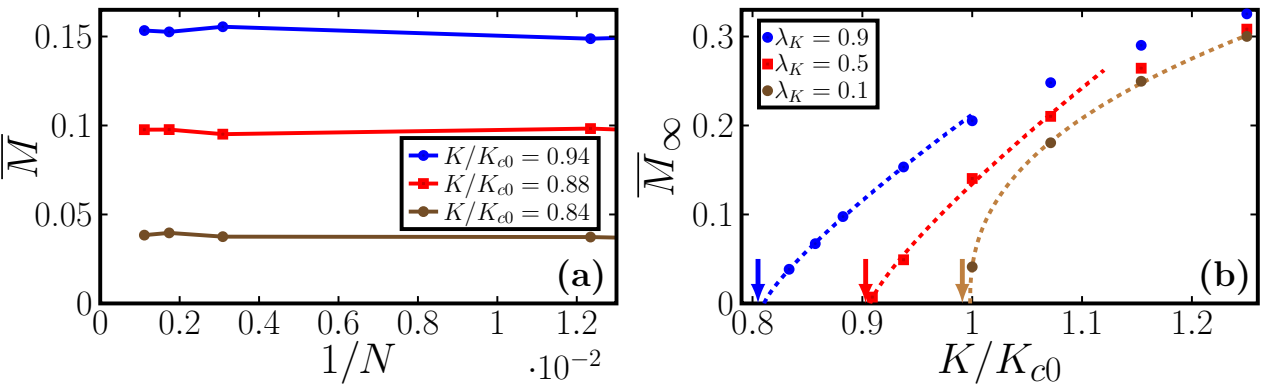

FIG. S2. (a) Finite-size scaling of mean values, $\overline{M}$, of the local magnetization for $\lambda_K = 0.9$; finite-size corrections are small. (b) Extrapolated mean of magnetization distribution, $\overline{M}_\infty$, as function of $K/K_{c0}$ for different $\lambda_K$. The respective value of critical point is consistent with that obtained from E-A order parameter (indicated by vertical arrows). The fitted critical exponent strongly depends on disorder.

can be located in multiple ways, as it is characterized by the disappearance of any triplon condensate. In Fig. 2 of the main text, we have employed the E-A order parameter of the spin glass and its finite-size scaling to obtain $K_c$.

Alternatively, one may use the distribution of local-moment amplitudes to locate the phase transition. Such distributions are shown in Fig. 3 of the main text and in Fig. S1, the latter for weaker disorder. Away from the transition, the distribution width scales essentially linearly with $\lambda_K$; their relative width increases strongly upon approaching the transition where eventually the distributions also become strongly non-Gaussian, Figs. 3 and S1(b). The mean value $\overline{M}$ can be extrapolated to infinite system size for fixed parameters, Fig. S2(a), with finite-size effects being negligible. These results for $\overline{M}_\infty$ enable an extrapolation towards the quantum critical point as shown in Fig. S2(b). Comparing this to the analysis of the E-A data in Fig. 2 of the main text, we notice that (i) the location of the transition is consistent between both approaches, (ii) the onset of the E-A order parameter $m_{sg}$ as function of $K$ is slower than that of $\overline{M}_\infty$, and (iii) the critical behavior deviates strongly from mean-field. (ii) and (iii) are related to the fact that order is strongly inhomogeneous near the transition; this is also illustrated in Fig. S3, i.e., there are relatively few sizeable-moment dimers forming islands embedded in a matrix of small-moment dimers. Since $m_{sg}$ involves a spatial average of squared moments, it grows more slowly than $\overline{M}$.

As explained in the main text, a Mott-glass phase is likely to intervene between the spin-glass and dimer phases. In the present approach, this would be signalled by the E-A order parameter going to zero after fluctuation corrections are taken into account. We leave such an analysis for future work.

We re-iterate that conventional long-range magnetic order is absent for any finite level of bond disorder [12]. However, the magnetic correlation length is exponentially large for small disorder, as bond disorder is a marginally relevant perturbation to the non-collinear ordered state in $d = 2$. As a result, finite-size simulations may effectively see an ordered state. This has been discussed extensively in Ref. 12. In our simulations we have also monitored the static spin structure factor, in particular the evolution of its peak at $\vec{Q} = (4\pi/3, 0, \pi)$ (equivalently, $(2\pi/3, \pm 2\pi/\sqrt{3}, \pi)$) corresponding to 120° long-range order. We show sample results in Fig. S4 for elevated disorder in the spin-glass phase; here the absence of LRO is indicated by the correlation length (as extracted from the peak width) extrapolating to a finite value as $N \to \infty$. In general, our results are consistent with the conclusions of Ref. 12.

## IV. SPIN GLASS: ENERGY LANDSCAPE AND CHIRALITIES

A spin glass is characterized by a rugged energy landscape, with many local minima at very low energies, separated by barriers. Our iterative algorithm to find the optimum product state is designed to find local minima, and we can therefore monitor their energy distribution. Glassiness is signalled by few hits at the lowest energy $E_0$ and many hits at different energies close to $E_0$. A non-glassy state instead leads to many hits at $E_0$ and few (if at all) hits at significantly higher energies.

Fig. S5 shows histograms of saddle-point energies in the spin-glass phase. A multitude of states within a window $\Delta E$ of $10^{-6}J$ above the ground-state energy $E_0$ are visible; if the system moves deeper into the glass phase, there is an increasing number of metastable states also at higher energies up to $\Delta E \sim 10^{-2}J$.

The results support the conclusion stated in the main text that glassiness is strong deep in the glass phase, $K \gg J$, but becomes weak near the transition to the paramagnet. This is consistent with the analysis of the vector chiralities and relative angles between neighboring spins, Fig. 4 of the main text, which show that correlations become progressively collinear upon approaching the transition, because the inhomogeneities in moment amplitude effectively reduce frustration. These inhomogeneities in moment amplitude can be nicely seen in the snapshots in Fig. S3: In panel (a) elementary triangles where all three dimers have a substantial moment amplitude are rare.

For completeness, we also show results for the scalar

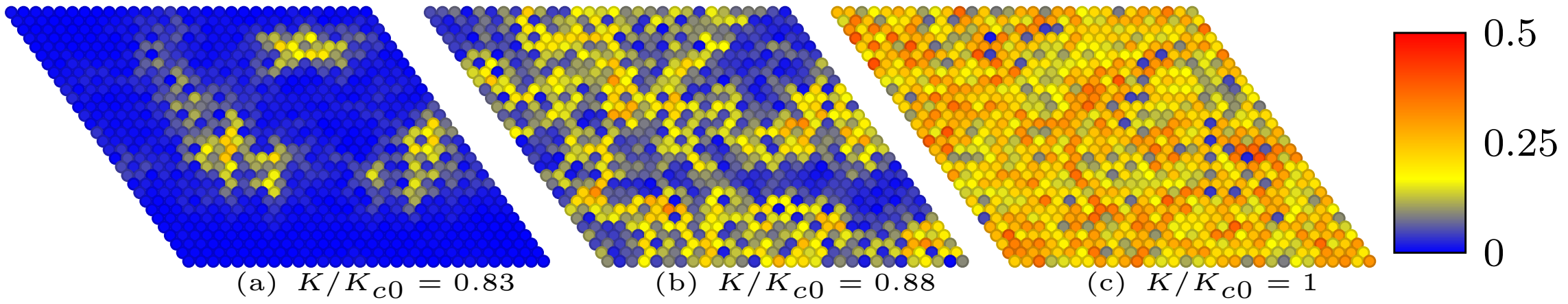


FIG. S3. Spatial distribution of local magnetization $M_{1i}$ for a fixed realization of intralayer box disorder, with each dot representing a dimer site of a $L = 30$ system. (a) $K/K_{c0} = 0.83$, (b) $K/K_{c0} = 0.88$, (c) $K/K_{c0} = 1$.

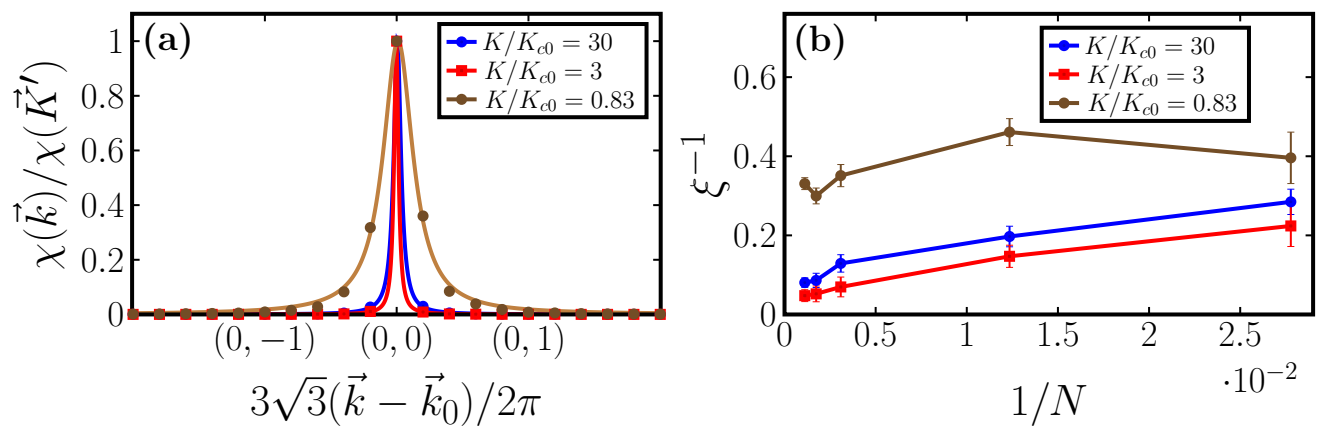


FIG. S4. (a) Normalized static spin structure factor for strong intralayer box disorder, $\lambda_K = 0.9$, and different $K/K_{c0}$ at fixed system size, together with Lorentzian fits near $\vec{Q}$. (b) Finite-size scaling of the inverse magnetic correlation length, extracted from the fits as shown in (a), indicating the absence of conventional long-range magnetic order.

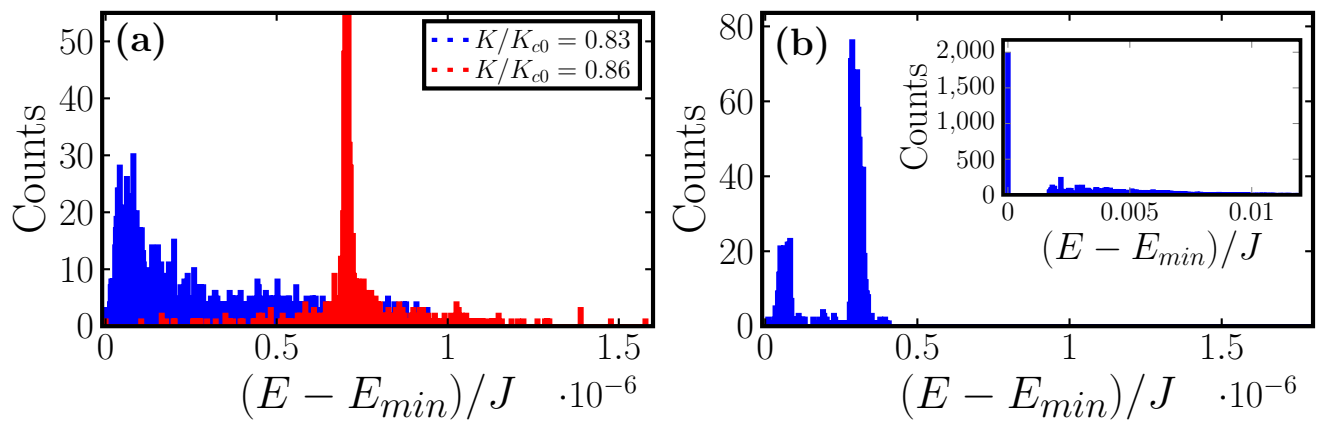


FIG. S5. Probability distribution of converged energies for a fixed disorder realization, obtained at the product-state level for intralayer box disorder with $\lambda_K = 0.9$. (a) $K/K_{c0} = 0.83, 0.86$, i.e., close to the glass transition, and 2000 random initializations. (b) $K/K_{c0} = 3.0$, i.e., deep in the spin-glass phase, and $10^4$ random initializations. The inset shows that metastable states also exist at significantly higher energies.

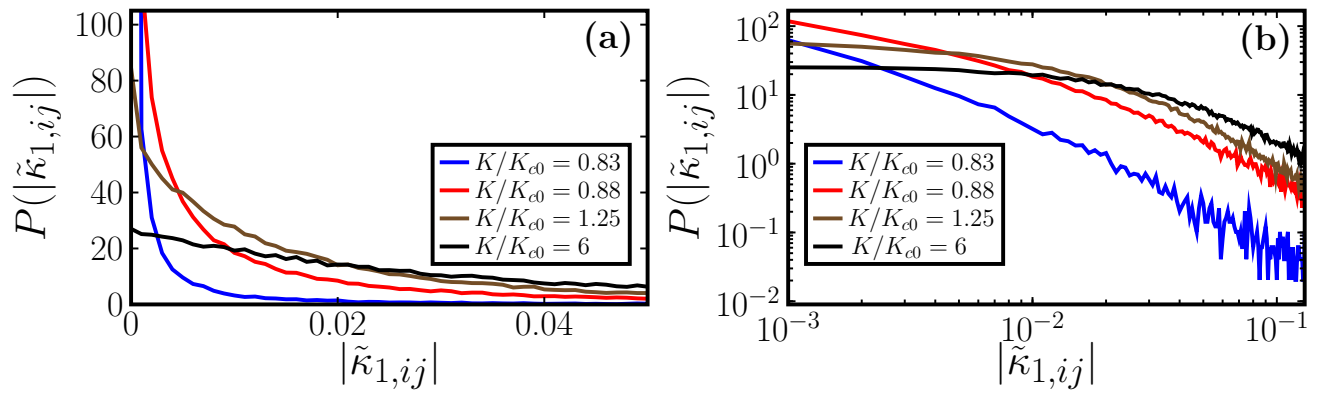


FIG. S6. Probability distribution of normalized scalar chirality $\tilde{\kappa}_{1,ijk} = \kappa_{1,ijk}/(|\vec{S}_{1i}||\vec{S}_{1j}||\vec{S}_{1k}|)$ for intralayer box disorder of strength $\lambda_K = 0.9$, indicating non-coplanar configurations. (a) Absolute value of chirality. (b) Same data on a log-log plot. The data cover regimes both deep in the spin-glass phase ($K/K_{c0} = 6$) as well as near criticality ($K/K_{c0} \lesssim 1$).

chirality $\kappa$, normalized to the moment sizes, in Fig. S6. These chiralities are non-zero in the spin-glass phase, fluctuating around zero. This indicates that the coplanar 120° state of the clean limit is rendered non-coplanar by disorder; these non-coplanarities are random and increase deep in the spin-glass phase.

## V. EVOLUTION OF SPIN EXCITATIONS WITH DISORDER

In this section, we show additional data for the dynamic spin susceptibility $\chi''(\vec{k}, \omega)$, computed as described in Sec. I C, which complement Fig. 5 of the main text. Fig. S7 illustrates the behavior of $\chi''(\vec{k}, \omega)$ with increasing bond disorder for different ratios of the intralayer-to-interlayer coupling.

Inside the ordered phase, the clean limit displays the bilayer-hybridized transverse Goldstone modes, with weight near momentum $K$ in the $k_z = \pi$ sector, Fig. S7(a1). We recall that the coplanar 120° order features two types of Goldstone modes, corresponding to in-plane and out-of-plane spin rotations, respectively, with different velocities. $\chi''(\vec{k}, \omega)$ also shows signatures of the longitudinal (Higgs) mode, whose weight decreases upon moving deeper into the ordered phase. With increasing disorder, the transverse modes at intermediate energy get broadened and loose intensity, while the low-energy modes remain relatively well defined, Fig. S7(b1-d1). In contrast, the Higgs-mode intensity disappears already for weak disorder, see Sec. VIII for further discussion.

At the clean-limit QPT, Fig. S7(b1), we see a soft triplon mode; we note that quantum critical broadening and higher-order continua are not part of the present approximation. Upon increasing the disorder level, the mode gets broadened, now in particular at low energies. We recall that increasing disorder effectively moves the system away from the transition and into the spin-glass phase; the low-energy excitations now become the Goldstone mode of the glass which itself is strongly disorder-broadened due to localization effects.

Finally, starting in the clean-limit dimer phase, Fig. S7(c1), we see that increasing disorder moves the system across the transition into the glass state, Fig. S7(c4). Correspondingly, we see a closing of the gap and the sub-

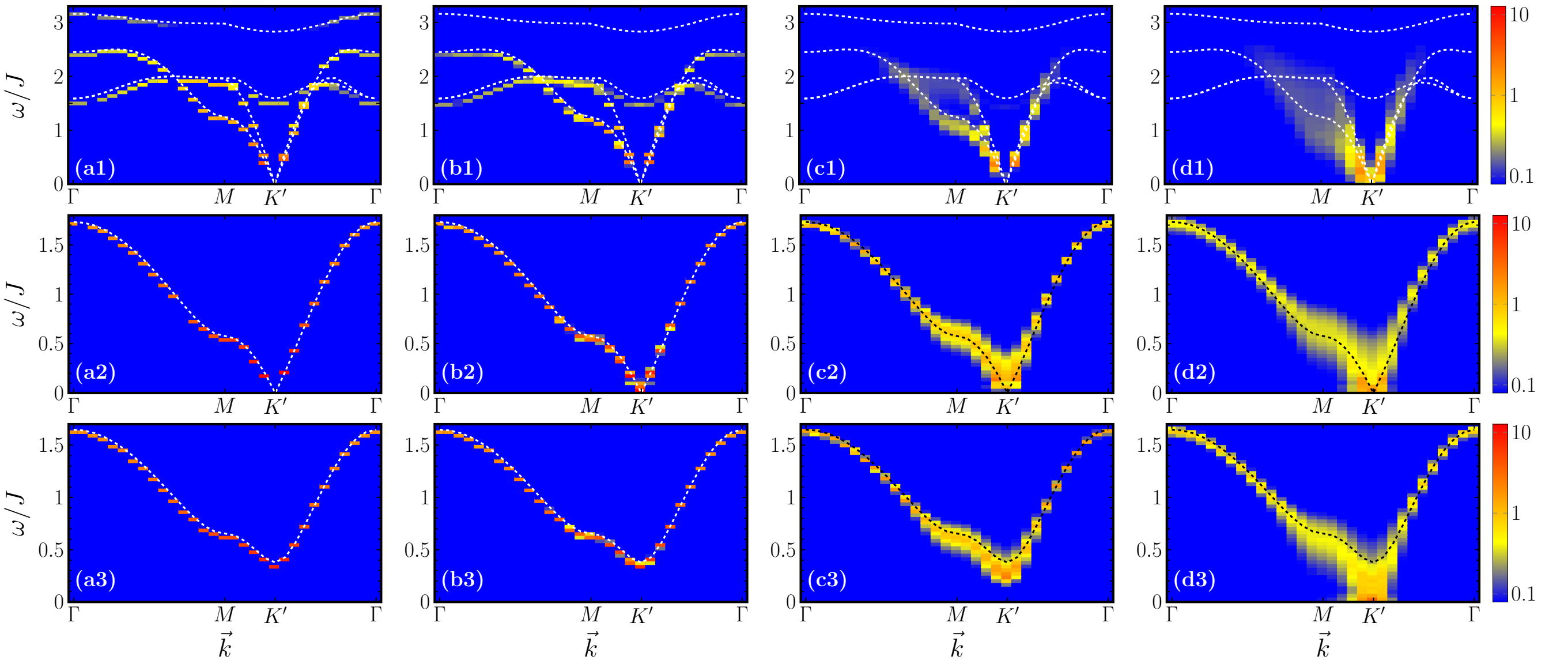


FIG. S7. Dynamical spin susceptibility $\chi''(\vec{k},\omega)$ as in Fig. 5 of the main text, here for different levels of intralayer bond disorder: (a) $\lambda_K = 0$, i.e., the clean limit, (b) $\lambda_K = 0.1$, (c) $\lambda_K = 0.5$. (d) $\lambda_K = 0.9$. The rows correspond to different coupling ratios: $K/K_{c0} = 3.0$ (top), $K/K_{c0} = 1.0$ (middle), and $K/K_{c0} = 0.86$ (bottom). Dashed curves show the clean-limit dispersion, with two magnons plus one Higgs mode in the top row, and degenerate triplon in the other rows.

sequent emergence of zero-energy modes with increasing disorder.

## VI. MODE LOCALIZATION

The inverse participation ratio (IPR) of the eigenmodes can be used to quantify their degree of spatial localization, with the general definition of IPR $= \sum_i |\psi_i|^4$ for a normalized wavefunction $|\psi\rangle$, and $\psi_i = \langle i|\psi\rangle$ where $i$ is a lattice site. For spatially extended states, the IPR scales with the system size as $1/N$, whereas it approaches a constant for localized states. For exponential localization, the IPR value is then a measure of the localization length, IPR $\propto 1/\xi^d$.

For the eigenvectors $\tau_n$ (S8) of the bosonic Bogoliubov transformation we define the IPR as

$$\mathrm{IPR}_n = \sum_{i\alpha} \left(|u_{ni\alpha}|^2 - |v_{ni\alpha}|^2\right)^2, \tag{S17}$$

note that $\sum_{i\alpha}(|u_{ni\alpha}|^2 - |v_{ni\alpha}|^2) = 1$.

Numerical results are shown in Fig. S8. For the present case of box disorder, localization happens primarily near the band edges, both in the dimer and glass phases. Modes are most strongly localized at low energies near the QPT into glass phase, Fig. S8(b,c): These are the Goldstone modes of the glass whose spatial structure follows the profile of the local magnetization. The latter is extremely inhomogeneous near the transition, Fig. S3(a), such that these modes display a sizeable amplitude on dilute sites only.

Notably, deep in the glass phase, Fig. S8(d), the elevated-energy modes appear all localized. These modes

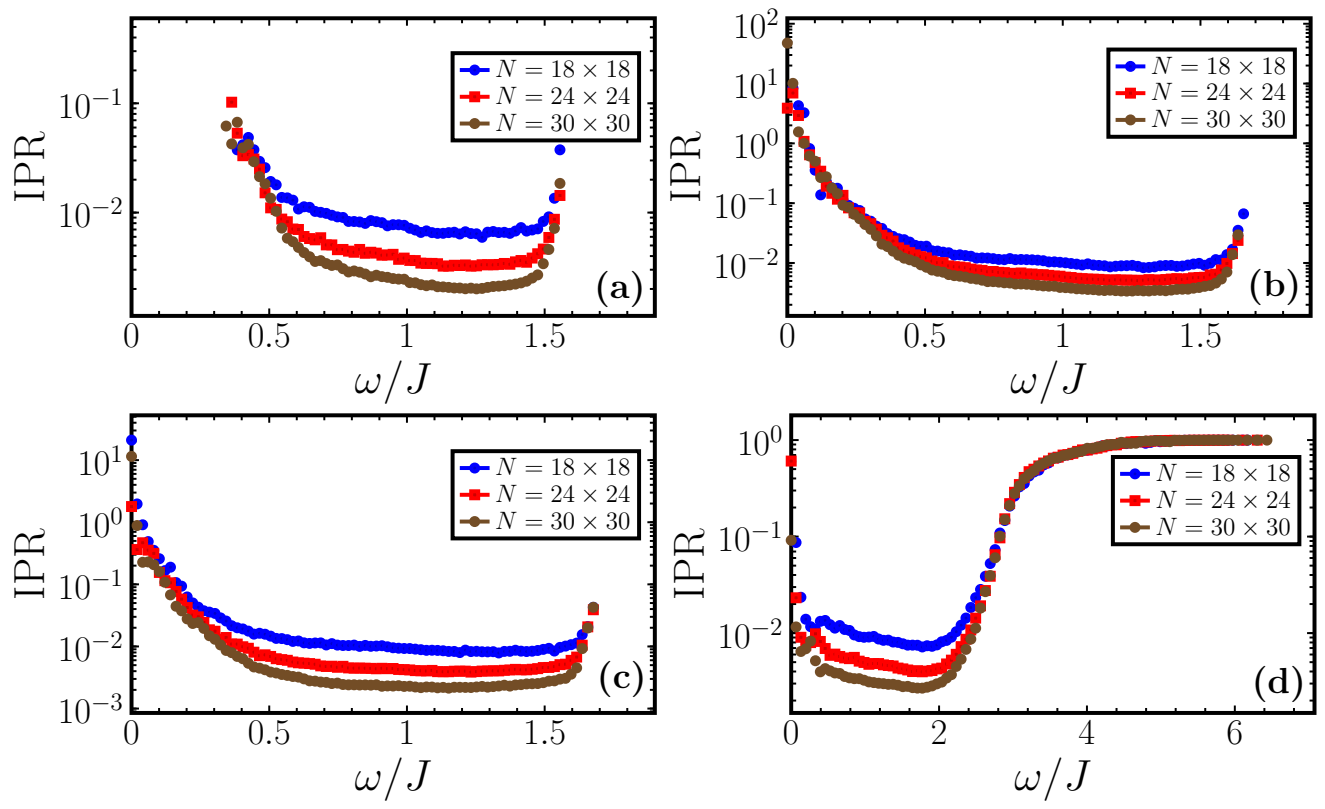


FIG. S8. Inverse participation ratio of excitation modes for intralayer box disorder with $\lambda_K = 0.9$, obtained for $L = 30$. (a) $K/K_{c0} = 0.67$, (b) $K/K_{c0} = 0.79$, (c) $K/K_{c0} = 0.83$, (d) $K/K_{c0} = 3.0$.

derive from amplitude excitations, see also Fig. S10 below. They are subject to large randomness due to the strongly inhomogeneous distribution of both staggered and uniform magnetization per dimer.

## VII. LOW-ENERGY EXCITATIONS

While the spectrum is gapped in the dimer phase, it is gapless throughout the glass phase due to the broken $SU(2)$ spin symmetry. We have computed the density of states (DOS) of the generalized triplon excitations, with results displayed in Fig. S9. In the clean-limit ordered

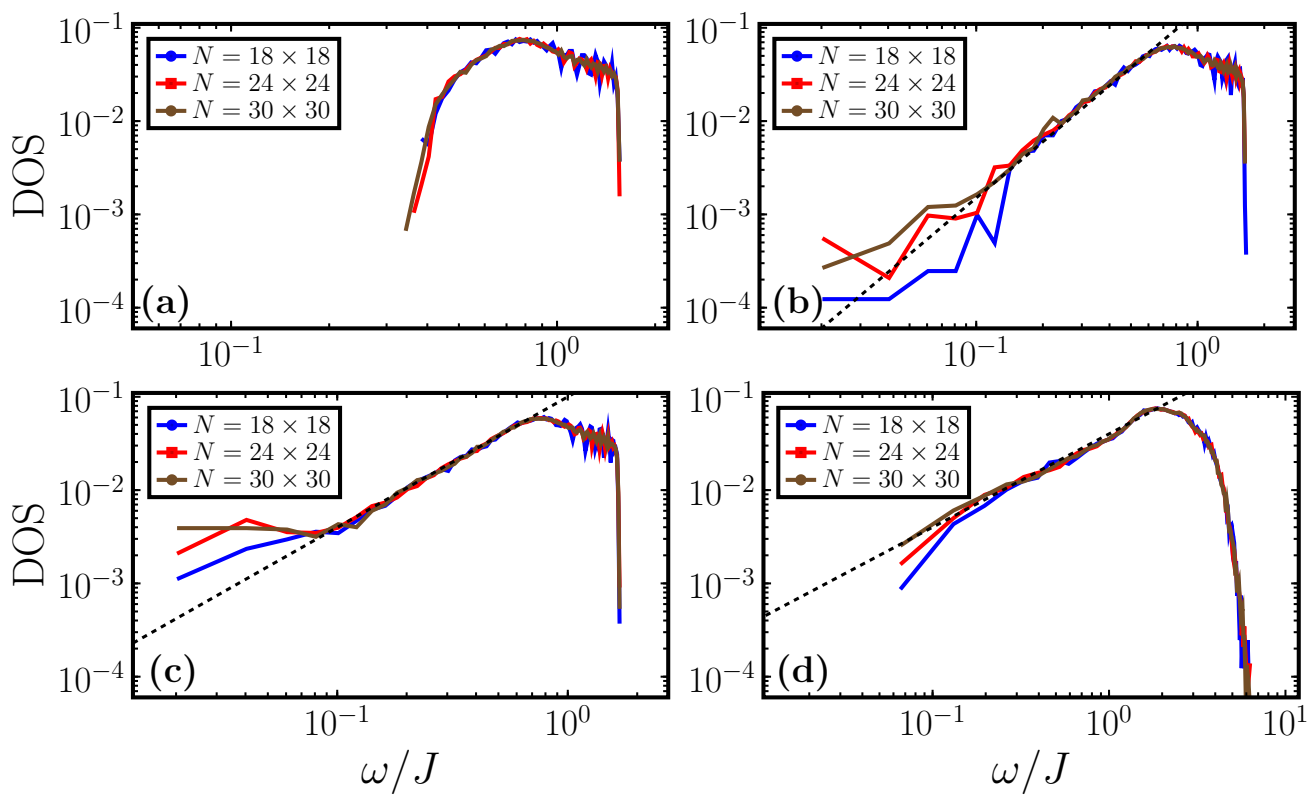


FIG. S9. Density of states of spin excitations, plotted on a log-log scale, for intralayer box disorder with $\lambda_K = 0.9$ and different system sizes. (a) $K/K_{c0} = 0.67$, (b) $K/K_{c0} = 0.79$, (c) $K/K_{c0} = 0.83$, (d) $K/K_{c0} = 3.0$. The dashed lines show power laws $\rho(\omega) \propto \omega^\kappa$ with exponent $\kappa = 2$ in (b), $\kappa = 1.4$ in (c) and $\kappa = 1$ in (d).

phase, the linearly dispersing Goldstone modes produce a DOS $\rho(\omega) \propto \omega$ (not shown). With quenched disorder included, we continue to observe $\rho(\omega) \propto \omega$ at the lowest accessible energies, although finite-size limitations render a precise statement difficult.

We note that a linear DOS scaling is consistent with the hydrodynamic theory of Halperin and Saslow [13] which predicts linearly dispersing low-energy modes in the Heisenberg spin glass. However, the agreement is likely fortuitous: As discussed in Ref. 14, the spin-wave modes of the two-dimensional spin glass are Anderson-localized (albeit perhaps weakly) in the independent-boson approximation (this is consistent with our IPR results described above), such that the hydrodynamic theory is not applicable. Interactions between modes, beyond the present approximation, can be expected to recover the hydrodynamic picture. Returning to our results in Fig. S9(d), we consider it therefore possible that the asymptotic low-energy behavior of the DOS within our approximation is different from $\rho(\omega) \propto \omega$, i.e., the latter is an intermediate-scale crossover only.

Near criticality, Fig. S9(b,c) we observe a larger exponent, $\rho(\omega) \propto \omega^2$, in an intermediate range of energies where our data are essentially free of finite-size effects. As above, determining the asymptotic low-energy behavior of $\rho(\omega)$ at criticality is hampered by finite-size effects.

## VIII. FATE OF HIGGS MODE

The dynamic bond-bond (or singlet) correlations allow to track the fate of the amplitude (or Higgs) mode [15–18]; recall that this mode is nicely visible in the spin-spin correlations only very close to the quantum phase transition. In Fig. S10 we show the disorder evolution of the bond susceptibility, $\chi_b''(\vec{q}, \omega)$. In the ordered phase, the well-defined Higgs mode, Fig. S10(a1), gets broadened by disorder, but also drastically looses intensity, Fig. S10(c1,d1). As explained in the main text, this is primarily because intralayer disorder breaks the $\mathbb{Z}_2$ symmetry of layer exchange, such that there is a sizeable and random *uniform* magnetization per dimer. As a result, the amplitude eigenmode acquires randomized phases relative to $B_q|0\rangle$, such that destructive interference reduces the mode intensity. Starting at the clean-limit critical point, Fig. S10(a2), increasing bond disorder moves the system into the glass phase. This generates some Higgs signal in $\chi_b''$, visible near $\vec{q} = \Gamma$, but with very small intensity only, Fig. S10(c2,d2). This becomes even further reduced for smaller $K/K_{c0}$, as shown in Fig. 5 of the main text.

We recall that multi-particle processes relevant for the Higgs mode, in particular its decay into two transverse excitations, are not part of the linear bond-operator approach employed here. This decay is known to be important deep inside the ordered phase where it leads to renormalization and – more importantly – significant damping of the Higgs mode [15–19]. In the present disordered situation, the Higgs-mode intensity is therefore suppressed by both the interference effect described above and two-particle decay.

## IX. SELECTED RESULTS FOR BIMODAL BOND DISORDER

While all results shown so far were for disorder distribution of uniform box shape, we have also studied the model with bimodal bond disorder. For intralayer bond disorder we randomly choose a fraction $p$ bonds out of the $6N$ intralayer bonds to which we assign a coupling strength different from the other bonds, such that

$$K_{m,ij} = \begin{cases} \alpha K\,, & \text{fraction } p \text{ bonds,} \\ K\,, & \text{fraction } (1-p) \text{ bonds.} \end{cases} \tag{S18}$$

Similarly, for the interlayer bond randomness, we choose $p'$ bonds out of the $N$ interlayer bonds and assign a coupling strength different from the others,

$$J_i = \begin{cases} \beta J\,, & \text{fraction } p' \text{ bonds,} \\ J\,, & \text{fraction } (1-p') \text{ bonds.} \end{cases} \tag{S19}$$

Hence, $p$ and $p'$ play the role of a defect concentration.

Many aspects of the results for bimodal disorder qualitatively agree with that obtained from box disorder. Quantitative differences in thermodynamics concern the shift and shape of the phase boundaries, as bimodal disorder with $\alpha < 1$ or $\beta > 1$ tends to stabilize the dimer paramagnet.

Significant differences naturally occur in the dynamic response, as bimodal disorder can lead to impurity bands whose energy lies outside that of the main excitation modes. An example for this is shown in Fig. S11 in the dimer phase: a fraction of weak interlayer bonds

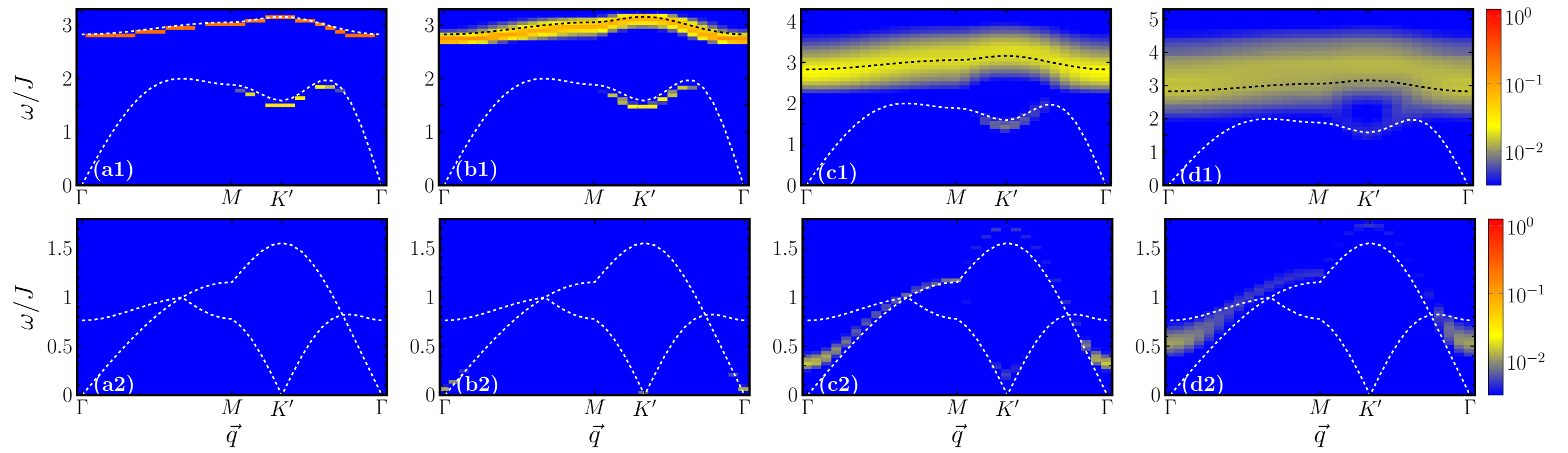


FIG. S10. Dynamical bond susceptibility $\chi''_b(\vec{q},\omega)$ as in Fig. 5 of the main text, here for different levels of intralayer bond disorder: (a) $\lambda_K = 0$, i.e., the clean limit, (b) $\lambda_K = 0.1$, (c) $\lambda_K = 0.5$. (d) $\lambda_K = 0.9$. The rows correspond to $K/K_{c0} = 3.0$ (top), $K/K_{c0} = 1.0$ (bottom). Dashed curves show the clean-limit dispersion for the Goldstone mode and the amplitude (Higgs) mode. In the clean limit, at $K/K_{c0} = 1.0$ the Higgs modes becomes soft.

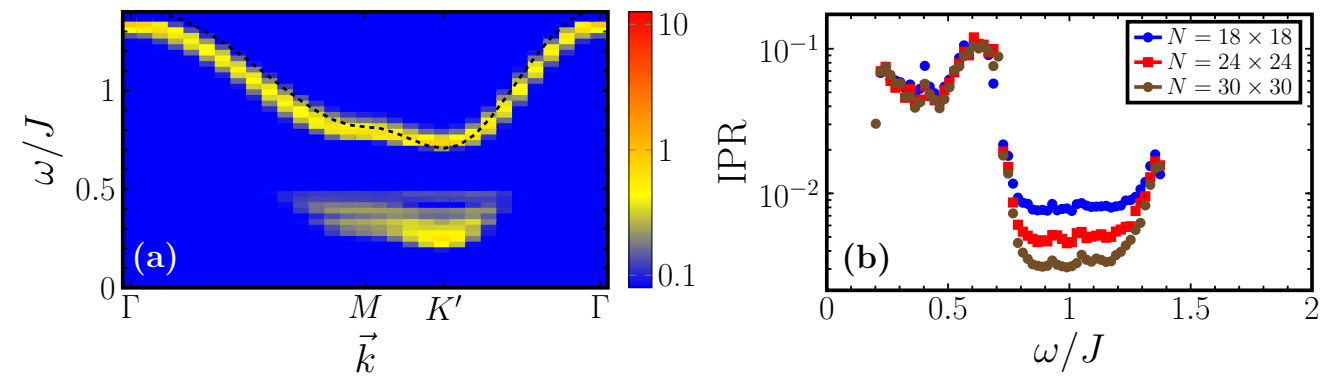


FIG. S11. Dynamical spin structure factor and IPR of excitation modes for bimodal interlayer bond disorder with $\beta = 0.5$, $p' = 0.3$, $K/K_{c0} = 0.5$.

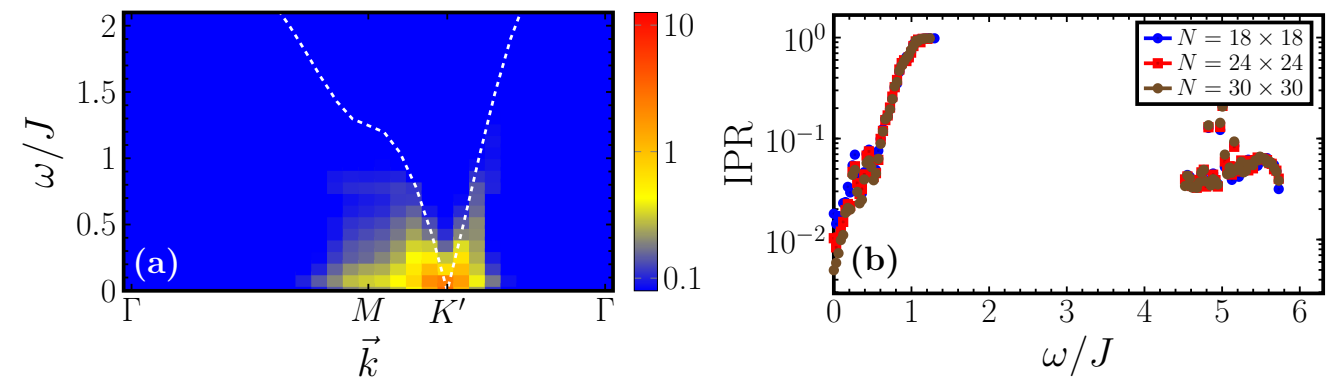


FIG. S12. Dynamical spin structure factor and IPR of excitation modes for bimodal interlayer bond disorder with $\beta = 15$, $p' = 0.5$, $K/K_{c0} = 3$.

($\beta < 1$) nucleates impurities which locally have a smaller triplon gap, leading to a sub-gap state whose main intensity is at $\mathbf{k} = \mathbf{Q}$, Fig. S11(a). The corresponding states are strongly localized, as indicated by the IPR re-

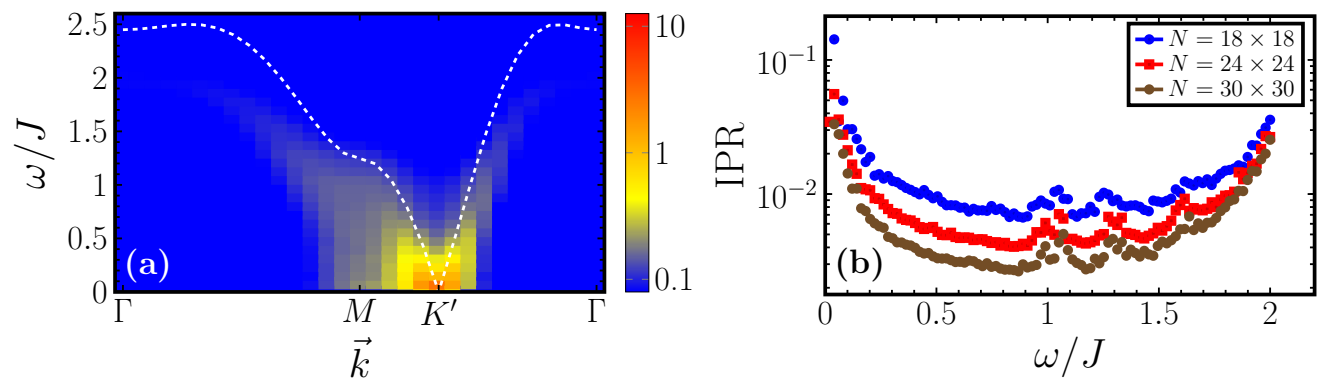


FIG. S13. Dynamical spin structure factor and IPR of excitation modes for bimodal intralayer bond disorder with $\alpha = 0.1$, $p = 0.5$, $K/K_{c0} = 3$.

sults in Fig. S11(b). Bimodal interlayer disorder can also mimic the physics of magnetic vacancies, as illustrated in Fig. S12. Here, half of the interlayer bonds are very strong, such that the corresponding dimers only weakly participate in the magnetic order. Thus, the system is effectively close to its percolation threshold, leading to spatial localization and strong momentum-space smearing of all low-energy modes.

Bimodal intralayer disorder can similarly lead to mode localization, but this happens primarily near the band edges, Fig. S13, as in the case of box disorder. The primary reason is that uncorrelated intralayer disorder is more efficiently averaged over, due to the large number of intralayer nearest neighbors, as compared to interlayer disorder.

---


[1] S. Sachdev and R. N. Bhatt, Bond-operator representation of quantum spins: Mean-field theory of frustrated quantum Heisenberg antiferromagnets, Phys. Rev. B **41**, 9323 (1990).

[2] T. Sommer, M. Vojta, and K. W. Becker, Magnetic properties and spin waves of bilayer magnets in a uniform field, Eur. Phys. J. B **23**, 329 (2001).

[3] M. Matsumoto, B. Normand, T. M. Rice, and M. Sigrist, Magnon dispersion in the field-induced magnetically ordered phase of $TlCuCl_3$, Phys. Rev. Lett. **89**, 077203 (2002).

[4] M. Vojta, Excitation spectra of disordered dimer magnets near quantum criticality, Phys. Rev. Lett. **111**, 097202 (2013).

[5] M. Puschmann, J. Crewse, J. A. Hoyos, and T. Vojta, Collective modes at a disordered quantum phase transition, Phys. Rev. Lett. **125**, 027002 (2020).

[6] S. Wessel and I. Milat, Phys. Rev. B **71**, 104427 (2005).

[7] V. N. Kotov, O. P. Sushkov, W. H. Zheng, and J. Oitmaa, Novel approach to description of spin-liquid phases in low-dimensional quantum antiferromagnets, Phys. Rev. Lett. **80**, 5790 (1998).

[8] D. G. Joshi, K. Coester, K. P. Schmidt, and M. Vojta, Nonlinear bond-operator theory and $1/d$ expansion for coupled-dimer magnets. I. Paramagnetic phase, Phys. Rev. B **91**, 094404 (2015).

[9] D. G. Joshi and M. Vojta, Nonlinear bond-operator theory and $1/d$ expansion for coupled-dimer magnets. II. Antiferromagnetic phase and quantum phase transition, Phys. Rev. B **91**, 094405 (2015).

[10] A. V. Syromyatnikov, Collective excitations in spin-1/2 magnets through bond-operator formalism designed both for paramagnetic and ordered phases, Phys. Rev. B **98**, 184421 (2018).

[11] S.-S. Zhang, Y. Kato, E. A. Ghioldi, L. O. Manuel, A. E. Trumper, and C. D. Batista, Large-N SU(4) Schwinger boson theory for coupled-dimer antiferromagnets, Phys. Rev. B **112**, 024410 (2025).

[12] S. Dey, E. C. Andrade, and M. Vojta, Destruction of long-range order in noncollinear two-dimensional antiferromagnets by random-bond disorder, Phys. Rev. B **101**, 020411 (2020).

[13] B. I. Halperin and W. M. Saslow, Hydrodynamic theory of spin waves in spin glasses and other systems with noncollinear spin orientations, Phys. Rev. B **16**, 2154 (1977).

[14] G. Bracci-Testasecca, J. Niedda, A. Coraggio, R. Moessner, and A. Scardicchio, Semiclassical picture of the Heisenberg spin glass in two dimensions: from weak localization to hydrodynamics (2026), arXiv:2603.22077.

[15] D. Podolsky, A. Auerbach, and D. P. Arovas, Visibility of the amplitude (Higgs) mode in condensed matter, Phys. Rev. B **84**, 174522 (2011).

[16] D. Podolsky and S. Sachdev, Spectral functions of the Higgs mode near two-dimensional quantum critical points, Phys. Rev. B **86**, 054508 (2012).

[17] M. Lohöfer, T. Coletta, D. G. Joshi, F. F. Assaad, M. Vojta, S. Wessel, and F. Mila, Dynamical structure factors and excitation modes of the bilayer Heisenberg model, Phys. Rev. B **92**, 245137 (2015).

[18] Y. Q. Qin, B. Normand, A. W. Sandvik, and Z. Y. Meng, Amplitude mode in three-dimensional dimerized antiferromagnets, Phys. Rev. Lett **118**, 147207 (2017).

[19] S. H. Do, H. Zhang, T. J. Williams, T. Hong, V. Ovidiu Garlea, J. A. Rodriguez-Rivera, T.-H. Jang, S. W. Cheong, J.-H. Park, C. D. Batista, and A. D. Christianson, Decay and renormalization of a longitudinal mode in a quasi-two-dimensional antiferromagnet, Nat. Commun. **12**, 5331 (2021).



\end{document}